\documentclass[aip, amsmath,amssymb, reprint]{revtex4-1}
\usepackage{graphicx}
\usepackage{dcolumn}
\usepackage{bm}
\usepackage{tabularx}
\usepackage[utf8]{inputenc}
\usepackage[T1]{fontenc}
\usepackage{mathptmx}
\usepackage{amsmath}
\usepackage{color}
\usepackage[colorlinks=true,linkcolor=red]{hyperref}
\newcommand{\fin}[1]{\color{black} #1\color{black}}
\newcommand{\mh}[1]{\color{black} #1\color{black}}
\begin{document}
\title{\fin{Time correlation functions for quantum systems: validating Bayesian approaches for harmonic oscillators and beyond}
}
\author{Vladislav Efremkin}
\email{vladislav.efremkin@univ-grenoble-alpes.fr}
\affiliation{ Univ. Grenoble Alpes, CNRS, LIPhy, 38000 Grenoble, France}
\author{Jean-Louis Barrat}
\email{jean-louis.barrat@univ-grenoble-alpes.fr}
\affiliation{
Univ. Grenoble Alpes, CNRS, LIPhy, 38000 Grenoble, France
}
\author{Stefano Mossa}
\affiliation{ CEA, IRIG-MEM, Univ. Grenoble Alpes, 38000 Grenoble, France}
\author{Markus Holzmann}
\affiliation{Univ. Grenoble Alpes, CNRS, LPMMC, 38000 Grenoble, France
}
\date{\today}
\begin{abstract}
The quantum harmonic oscillator is the fundamental building block to compute thermal properties of virtually any dielectric crystal at low temperatures in terms of phonons, extended further to cases with anharmonic couplings, or even disordered solids. In general, Path Integral Monte Carlo (PIMC) or Molecular Dynamics (PIMD) methods are powerful tools to determine stochastically thermodynamic quantities without systematic bias, not relying on perturbative schemes. Addressing transport properties, for instance calculating thermal conductivity from PIMC, however, is substantially more difficult. Although correlation functions of current operators can be determined by PIMC from analytic continuation on the imaginary-time axis, Bayesian methods are usually employed for the numerical inversion back to real-time response functions. This task not only strongly relies on the accuracy of the PIMC data, but also introduces noticeable dependence on the model used for the inversion. Here, we address both difficulties with care. In particular, we first devise improved estimators for current correlations which substantially reduce the variance of the PIMC data. Next, we provide a neat statistical approach to the inversion problem, blending into a fresh workflow the classical stochastic maximum entropy method together with recent notions borrowed from statistical learning theory. We test our ideas on a single harmonic oscillator and a collection of oscillators with a continuous distribution of frequencies, and provide indications of the performance of our method in the case of a particle in a double well potential. This work establishes solid grounds for an unbiased, fully quantum mechanical calculation of transport properties in solids.
\end{abstract}
\maketitle
\section{Introduction}
\label{sec:Intro}
Atomic nuclei in condensed phases behave, in many cases, as quantum objects. For instance, Nuclear Quantum Effects are responsible for the {\em heat capacity problem}, i.e., the deviation from the classical Dulong and Petit law for the heat capacity of solids at low temperatures. The solution of this issue eventually led to the development of the harmonic theory of solids, an accurate quantum theory that lets us to compute their thermal properties at temperatures lower than the Debye temperature, and can be corrected to account for anharmonic effects~\cite{BornHuang,AshcroftMermin}. By reducing the description of an insulating solid to a set of independent harmonic oscillators, the {\em phonons}, weakly interacting through anharmonic couplings, this theory also provides a framework for the computation of transport properties, in particular of heat conductivity. In contrast to the very high accuracy that can be achieved for thermal properties, however, the computation of transport features is sensibly more delicate and often requires ad hoc approximations for the lifetime of phonons, which is limited by phonon-phonon scattering processes and the presence of defects. 
The general framework of the harmonic theory of solids, originally developed for crystals, can be adapted to {\em disordered solids}. This is at the expenses of employing a numerical approach to characterize the harmonic eigenmodes, that replace phonons and are no longer determined by symmetries. Again, this procedure can be efficiently employed to determine thermal properties, while its application to transport is much more limited. Very often these properties are indeed calculated via classical statistical mechanics approaches (based on classical Molecular Dynamics simulations), whose results are next empirically corrected to account for quantum effects (see, among others, ~\cite{mizuno2016relation}). We also note that, in systems (ordered or disordered) involving light nuclei (e.g., hydrogen in solid ice), the large wavelength associated with light atoms makes the harmonic approximation itself  inappropriate. Therefore, an exact calculation should in general be considered even for thermal properties, or for the determination of phase boundaries~\cite{Bronstein2016}.

The harmonic theory of crystalline solids undoubtedly constitutes a remarkable achievement, as many results can be obtained based on an almost fully analytical approach. However, the above limitations in computing transport properties or in applying the theory to disordered structures, point to the necessity of numerical approaches. It would therefore be highly desirable to develop a numerical methodology that could fully take into account the quantum nature of atomic nuclei, allowing us to determine without approximations both thermal and transport properties of any insulating solid. 

When interested in thermal properties, an exact numerical method that encompasses all quantum aspects and is valid at any temperature, independently of the strength of anharmonic effects, involves the path integral representation of the partition function~\cite{Barker1979,Chandler1981,Herman1982,Pollock1984}. In the absence of exchange effects (a reasonable hypothesis in most common solids), the determination of thermodynamic properties at the inverse temperature $\beta =(k_B T)^{-1}$  involves the sampling of an equivalent system where each quantum particle is replaced by a discretized "path" consisting of $M$ "imaginary time slices". The method becomes exact in the limit of large $M$, and the sampling of $N$ quantum degrees  of freedom at temperature $T$ turns out to be equivalent to that of $N\times M$ classical degrees of freedom at temperature $M\times T$. This sampling can be achieved efficiently using Monte Carlo or Molecular Dynamics methods, leading to the PIMC and PIMD methods, \mh{respectively \cite{tuckerman2010statistical}}.

Computation of transport properties is more problematic. The standard Green and Kubo statistical mechanics approach to transport coefficients~\cite{Green1952,Kubo1957,Luttinger1964}, obtains the heat conductivity tensor $\kappa$ in a system of volume $V$ at temperature $T$ from a time correlation function of the energy current operator $\bf{J}$ as,
\begin{equation}
\kappa_{\alpha \beta } = \frac{1}{Vk_BT^2}\int_0^\infty dt \langle J_\alpha(t)J_\beta(0)\rangle.
\label{eq:k_ab}
\end{equation}
Unfortunately, the path integral method provides directly  static (time-independent) quantities only. A possible solution to this problem has been identified long ago~\cite{Thirumalai1983}, by noting that the PIMC approach can rather supply the analytical continuation of the correlation functions on the {\em imaginary time} axis, simply by computing the correlation between two imaginary time slices along the path. The power spectrum, $S_{AB}(\omega)$, of a real time correlator, $C_{AB}$, between two operators $A$ and $B$ can then be obtained in an apparently straightforward manner by using the identity,
\begin{equation}
C_{AB}(i\tau) = \int_0^\infty d\omega \left[S_{AB}(\omega) e^{-\hbar\omega\tau} + S_{BA}(\omega)e^{-\hbar\omega(\beta-\tau)}\right].
\label{eq:inversion}
\end{equation}

While Eq.~(\ref{eq:inversion}) in principle allows one to obtain $S$ based on the data for $C(i\tau)$, with $\tau$ in $[0,\beta]$, it is well known that the inversion problem is ill-posed, in the sense that determining $S$ with high precision is an extremely difficult task, even if $C$ is known with excellent accuracy. For this reason, the approach pioneered by a few groups in the eighties within the framework of path integral calculations did not spread widely. \mh{As an alternative, real-time correlation functions of the centroid, or ring-polymer dynamics computed within PIMD, often provide accurate spectra~\cite{Perez2009}. They lack, nonetheless, a rigorous justification of the underlying semi-classical approximation for obtaining low temperature transport properties, and we will not go this route in what follows.}

 Many recent studies  obtained  in various fields~\cite{PhysRevB.57.10287,Bertaina2017,LEVY2017149,PhysRevB.95.014102,PhysRevB.98.134509}, however, indicate that the present computing capabilities should by now allow us to carry out \mh{the original program of analytical continuation from imaginary to real time spectra} satisfactorily, by addressing the two major (and related) difficulties: {\em i)} to obtain with high accuracy the imaginary time correlation, in particular for current operators which suffer from the well known issue of diverging variance~\cite{Herman1982} in the limit of large $M$; and {\em ii)} to solve the ill-posed problem of extracting the frequency spectrum from the imaginary time correlation functions. 

Here we address these two issues based on numerical and analytical calculations of very simple examples, namely a single harmonic oscillator or an ensemble of oscillators with a continuum distribution of frequencies. The interest of this choice is twofold. First, due to its simplicity, we can obtain exact analytical expressions for most quantities of interest, including all time dependent correlations and exact expressions for the discretized path integrals. The availability of these expressions enables a precise control of the different sources of error, which can be both of statistical origin or associated with the discretization itself. Second, the harmonic oscillator is at the heart of the harmonic theory of solids, the natural starting point for any calculation of transport in insulating solids. Completely controlling this case is, therefore, crucial for any serious step forward in this direction. \mh{We conclude the study by providing promising indications of the performance of our method when applied to the strongly anharmonic case of a particle evolving in a double well potential.}

The manuscript is organized as follows: in Sect.~\ref{sec:PIMC} we introduce the general formalism of the path integral and imaginary time correlations, while in Sect.~\ref{sec:inversion} we present the procedure that we have developed to cope with the inversion problem. In Sect.~\ref{sec:estimators} we next describe a new approach that circumvents the issue of the diverging variance for current-current correlators. Finally, in Sects~\ref{sec:case1} and~\ref{sec:case2} we illustrate the application of these methods to a single harmonic oscillator, followed by the case of a collection of oscillators with a continuum distribution of frequencies, mimicking the density of states of a crystalline solid. In Sect.~\ref{sec:conclusion} we draw our conclusions.
\section{\label{sec:PIMC}The path integral formalism for time correlations}

The path integral Monte Carlo method provides a numerically exact route to the evaluation of thermodynamic properties of quantum systems at finite temperature, $T$. If we consider, for simplicity, a system described by a single degree of freedom $X$ of mass $m$, with Hamiltonian $ \hat{H} = \hat{P}^2/2m + U(\hat{X})$, the average value of an observable $\hat{A}$ is 
\begin{equation}
\langle\hat{A} \rangle =\frac{1}{Z(\beta)} \text{Tr} [\hat{A}\, e^{-\beta \hat{H} }  ],
\end{equation}
where $Z(\beta)=\text{Tr}[e^{-\beta\hat{H}}]$. In the PIMC approach, the trace is evaluated by expressing the density operator as $e^{-\beta\hat{H}}= (e^{-\beta\hat{H}/M})^M$. In the position representation $\vert X\rangle$, and using the notation $\rho(X,Y,\tau) = \langle X \vert e^{-\tau\hat{H}}  \vert   Y \rangle $, we can write
\begin{multline}
\langle  \hat{A} \rangle  = \frac{1}{Z(\beta)}
\int dX_0\ldots dX_M \\
\langle X_0 \vert \hat{A} \vert   X_1 \rangle \rho(X_1,X_2,\beta/M)\ldots\rho(X_M,X_0,\beta/M).
\label{eq:average1}
\end{multline}
If an expression for \fin{$\rho(X,Y,\beta/M)\;$} is known, the observable can be evaluated by sampling the "path" $\{X_0\ldots X_M\}$ with a statistical weight proportional to $\rho(X_1,X_2,\beta/M)\ldots\rho(X_M,X_0,\beta/M)$. As the matrix element $\langle X_0 \vert \hat{A} \vert X_1 \rangle$ of a {\em local} operator $\hat{A}$ involves in general a term $\delta(X_0-X_1)$, the sampling is actually performed over a closed path of $M$ points. In the following we will repeatedly consider the "primitive" approximation, based on the factorization of the kinetic and potential parts of the density operator and valid in the limit of small $\tau$\cite{Chandler1981},
\begin{equation}
\rho_p(X,Y,\tau) \simeq \sqrt{\frac{m}{2\pi\hbar^2\tau}}\exp\left\{-m\frac{ (X-Y)^2}{2\hbar^2\tau} -\frac{\tau}{2}\left[U(X)+U(Y)\right]\right\}.
\label{eq:primitive-approx}
\end{equation}
This simplified expression can be replaced by a more accurate one if needed, and if the exact value of $\rho$ is known, as it is the case for the harmonic oscillator, the latter can be used to sample the path  more efficiently~\cite{feynman1998statistical}.

Here, we are interested in equilibrium time correlation functions that determine the linear response properties of the system. A time correlation involving the observables $A$ at time $t$ and $B$ at time $t=0$ is the equilibrium average of the product of the operators $\hat{A}(t)= e^{itH/\hbar}\hat{A} e^{-itH/\hbar}$, and $\hat{B}(0)=\hat{B}$, which we can write as,
\begin{equation}
C_{AB}(t/\hbar) = \langle\hat{A}(t)\hat{B}(0)\rangle = \frac{1}{Z(\beta)}\text{Tr}[\hat{A}(t)\hat{B}(0)e^{-\beta\hat{H}}].
\end{equation}
Obviously, the splitting method could be applied to the operators $\exp(it\hat{H}/\hbar)$. Unfortunately, the statistical weight associated with the resulting path is imaginary, and therefore it is not suitable for usual sampling methods. If, however, the real time $t$ is replaced by an imaginary time $t=i\tau \hbar$, we can write,
\begin{multline}
C_{AB}(i\tau) =  \frac{1}{Z(\beta)}\text{Tr}  [\hat{A}e^{-\tau\hat{H}}\hat{B}e^{-(\beta-\tau)\hat{H}}]  \\
=\frac{1}{Z(\beta)} \int dX dX'dY dY' \\
\langle X \vert \hat{A} \vert   X' \rangle
\rho(X',Y,\tau)  \langle Y \vert \hat{B} \vert   Y' \rangle
\rho(Y',X,\beta -\tau),
\end{multline}
which is defined for $0 \le \tau \le  \beta$, and verifies $ C_{AB}(i\tau)=  C_{BA}(i(\beta-\tau))$. 

Partitioning again the interval $[0,\beta]$ into $M$ slices of width $\Delta\tau= \beta/M$, the correlation function can be sampled for discrete values of $\tau$ of the form $\tau_k=k\Delta\tau$, with $k=0\ldots M-1$, at a computational cost that is similar to that needed to calculate the thermodynamic observables of Eq.~(\ref{eq:average1}), obtaining
\begin{multline}
 C_{AB}(i\tau_k)= 
 \frac{1}{Z(\beta)}
  \int dXdY dX_1... dX_M \langle X \vert \hat{A} \vert   X_1 \rangle \rho(X_1,X_2,\Delta \tau)...\\ \rho(X_{k-1},X_k,\Delta\tau)
  \langle X_k \vert \hat{B} \vert   Y  \rangle
  \rho(Y,X_{k+1},\Delta\tau)...\rho(X_M,X,\Delta\tau).
\end{multline}
As in Eq.~(\ref{eq:average1}), here the sampling must be performed over the $\{X_1\ldots X_M\}$ coordinates of the path, the $X$ and $Y$ variables being eliminated by the $\delta$-functions contained in the matrix elements of $\hat{A}$ and $\hat{B}$.

\section{A statistical approach to the inversion problem}
\label{sec:inversion}
Once the imaginary time correlations, denoted by $C(\tau)$ from now on,  have been obtained for a set of $M$ discrete values $\{\tau_0...\tau_{M-1}\}$ in the interval $[0,\beta]$, the real time correlation functions relevant to describe the system physical response can, in principle, be obtained by inverting Eq.~(\ref{eq:inversion}). This is common to many studies of quantum systems, and generally described as the "analytical continuation" procedure. It is, however, ill-posed, in the sense that if the spectrum $S(\omega)$~\footnote{In this paragraph we drop the $AB$ subscripts in Eq.~(\ref{eq:inversion})} is described by a set of parameters (such as the values of $S$ on a discrete $\omega$-grid, or the coefficients of an expansion in terms of some basis set), and the $C(\tau_k)$ are affected by statistical errors, a very large number of solutions for $S$ compatible with the original data will be found.

This topic is the subject of a vast literature, and it is fair to conclude that no single method emerges as a preferred solution. Generally speaking, most current solutions employ some particular version of a "maximum entropy" approach~\cite{JARRELL1996,Boninsegni1996}. \mh{In the context of PIMC and to obtain real time data in combination with real time approximate methods, this procedure was used for instance in Refs.~\cite{Krilov1999,Krilov2001,Habershon2007}}. The spectral function, $S_{ME}$, is therefore obtained as an average over the possible $S(\omega)$'s (defined by some finite set of parameters), weighted by the probability that they are the exact model given the data set $(\textbf{C},\sigma^2)$,
\begin{equation}
S(\omega)_{ME} = \int \mathcal{D}S \ p(S|\text{C},\sigma^2)S(\omega).
\label{eq:sme1}
\end{equation}
Here $\mathcal{D}S$ indicates the phase space element associated with the parametrization of $S(\omega)$, $\textbf{C} = (C(\tau_1), C(\tau_2), \dots, C(\tau_M))^{\dagger} \equiv (C_1, C_2, \dots C_M)^{\dagger}$ is a line vector that contains the data points, and $\sigma^2$ describes the statistical uncertainty of these data in the form of a covariance matrix. By using the Bayes formula, one has,
\begin{equation}
p(S|\text{C},\sigma^2) = \frac{p(\textbf{C},\sigma^2|S)}{p(\textbf{C}, \sigma^2)}p(S),
\end{equation}
\mh{where $p(S)$ encompasses any prior information on the spectrum.} Making the assumption of  Gaussian statistics for the likelihood we can write,
\begin{equation}
p(\textbf{C}|S,\sigma) \propto e^{-\frac{1}{2}(\textbf{C} - \textbf{C}[S])(\sigma^2)^{-1} (\textbf{C} - \textbf{C}[S])} = e^{-\frac{1}{2}\chi^2[S]}\label{likelihood},
\end{equation}
which we can interpret as the definition of $\chi^2[S]$. Here $\textbf{C}[S]$ is the expression of the vector $C$, obtained by inserting a known spectrum $S$ into the r.h.s. of Eq.~(\ref{eq:inversion}) and computing the resulting $M$ correlation values. In the case of a spectrum defined by the amplitudes $A(\omega_p)$ for a set of $N_\omega$ discrete frequencies on  a regular grid, using Eq.~(\ref{eq:inversion}) we obtain, 
\begin{equation}
\tilde{C}[S](\tau_\alpha) = \sum_{p=1}^{N_\omega} A(\omega_p) \left( e^{-\hbar\omega_p \tau_\alpha} + e^{-\hbar(\beta - \tau_\alpha)\omega_p}\right) \label{eq::correlation fit}.
\end{equation}

In traditional maximum entropy methods, Eq.~(\ref{eq:sme1}) is solved at the saddle point level, by minimizing the functional $\mathcal{F}=\frac{1}{2}\chi^2[S] - H[S]$. Here, $H[S]$ is an entropic functional, which assigns a penalty to irregular solutions that would lead to an overfitting of the statistical errors contained in the data. For a positive spectrum, $H[S]$ is usually chosen as the associated Shannon entropy, with a coefficient controlling the strength of the regularisation. 
In this work we employ the so-called "stochastic analytical inference" or "stochastic maximum entropy"~\cite{Fuchs2010} method, where Eq.~(\ref{eq:sme1}) is sampled by Monte-Carlo methods over $\mathcal{D}S$, which can be constrained to positive values of $S$ through the prior probability $p(S)$.  \mh{(In the case studies described below we have employed a flat prior.)} The term $\frac{1}{2}\chi^2[S]$ can hence be considered as an effective energy functional, and the method can be refined by introducing an additional parameter in the form of an effective inverse temperature $\Theta$ as,
\begin{equation}
S(\omega,\Theta)_{ME}= {Z(\Theta)^{-1}}\int \mathcal{D}S \   S(\omega) e^{-\frac{1}{2}\Theta \chi^2[S]}.
\label{eq:sme2}
\end{equation}
Here the normalization $Z(\Theta)= 1 / \exp{\{\Theta F(\Theta)\}}$ is an effective partition function. Note that the traditional maximum entropy approach corresponds to a mean field version of Eq.~(\ref{eq:sme2}), where one uses as an estimate of the spectrum the minimum of the mean field free energy $F_{MF}(\theta)= \frac{1}{2}\chi^2[S]- \Theta^{-1} H[S] $. In view of the following analysis, we make the simplifying assumption of uncorrelated data points, so that the covariance matrix is diagonal. As a result, we can write the energy functional $\chi^2[S]$ in the form, 
\begin{equation}
\chi^2 = \sum_{\alpha=0}^{M-1}\frac{[C(\tau_\alpha) - \tilde{C}[S](\tau_\alpha)]^2}{\sigma^2(\tau_\alpha)},
\label{eq:chi2}
\end{equation}
with $\sigma^2(\tau_\alpha)$ the statistical uncertainty on the data point $\alpha$. Several arguments~\cite{Fuchs2010} have been evoked for fixing $\Theta=1$. In contrast, in~\cite{Fuchs2010} it has been proposed to pick for $\Theta$ the value $\Theta^*$ that maximizes $Z(\Theta)$, which is argued to also maximize the posterior probability $P(\theta | C)$. This possibility, which corresponds to a balance between energy and entropy dominated solutions, requires however a full free energy calculation. \mh{We further note that increasing the value of  $\Theta$ is effectively equivalent to rescaling the uncertainties on the data points, a procedure that may lead to overfitting. The corresponding effect on our validation procedure is discussed in Sect.~\ref{subsec:inversion}.} 

At variance with these proposals, we optimize the value of $\Theta$ employing the following strategy. An initial data set, $C(\tau_\alpha)$, is generated with known statistical uncertainty $\sigma^2(\tau_\alpha)$ by using, for instance, a path integral simulation of the considered model. In cases were $C(\tau)$ is known analytically, synthetic data could also be generated from the exact solution, and introducing a controlled uncertainty. Starting from these data, the spectrum $S_{ME}(\Theta)$, described by $P$ degrees of freedom  $A(\omega_p)$, is obtained through a Monte-Carlo sampling of Eq.~(\ref{eq:sme2}) for a given value of $\Theta$. Note that a well converged Monte-Carlo average will lead to a spectrum $S_{ME}(\Theta)$ with an associated $\chi^2\sim \mathcal{O} (M\epsilon)$, where $\epsilon$ is a residual error, while the average $\langle \chi^2 \rangle \sim\mathcal{O}(M\epsilon+P/\Theta)$. We denote $\bar{C}_\Theta(\tau_\alpha)$ the correlation function associated with this average spectrum.

In order to determine the optimal choice of $\Theta$, therefore discriminating among different models for $S(\omega)$ (e.g., different finite discretizations on an $\omega$-grid), we combine the maximum entropy approach with a validation procedure borrowed from the  statistical learning theory~\cite{MEHTA20191}. We, therefore, generate $P'$ new sets of validation data, $C_{\mathrm{val}, i}(\tau_\alpha)$ ($i=1,\ldots, P'$), by using the same technique (even not necessarily with the same accuracy) that we use to produce the original data set, and determine the associated,
\begin{equation}
\chi^2_{\mathrm{val}} = \frac{1}{P'}\sum_{i=1}^{P'} \sum_{\alpha=0}^{M-1}[\bar{C}_\Theta(\tau_\alpha) - C_{\mathrm{val},i}(\tau_\alpha)]^2 .
\label{eq::chi2 validation}
\end{equation}
Interestingly, this can be interpreted as a measure of the difference between the estimate $\bar{C}_\Theta(\tau_\alpha)$ and the exact correlation function, denoted by ${C}_{\mathrm {exact}}(\tau_\alpha)$. Indeed, by writing 
\begin{equation}
 \chi^2_{\mathrm{val}}= \frac{1}{P'} \sum_{i=1}^{P'} \sum_{\alpha=0}^{M-1} [\bar{C}_\Theta(\tau_\alpha) - {C}_{\mathrm{exact}}(\tau_\alpha) + {C}_{\mathrm{exact}}(\tau_\alpha) - C_{val,i}(\tau_\alpha)]^2 ,
\end{equation}
in the limit of large $P'$ and assuming that the average over the validation data returns the exact correlation function, we obtain
\begin{equation}
 \chi^2_{\mathrm{val}}=  \sum_{\alpha=0}^{M-1} [\bar{C}_\Theta(\tau_\alpha) - {C}_{\mathrm{exact}}(\tau_\alpha)]^2 + \sum_{\alpha=0}^{M-1} \sigma^2_{\mathrm{val}}(\tau_\alpha).
 \label{eq::chi2_valid2}
\end{equation}
Here, the first term is the distance of the estimate to the exact data, while the second is the variance of the validation data \fin{leading to a background value $\chi^2_0$ independent of $\Theta$ (or any other parameter entering the model description).} The choice of $\Theta$ will therefore be eventually dictated by the behavior of the first term.
\section{\label{sec:estimators}Improved estimators for current correlations}
The computation of transport coefficients typically implies correlation functions involving the momentum operator, a prototypical one being $C_{pp}(\tau) = \langle p(\tau) p(0) \rangle $. In the path integral approach and within the primitive approximation of Eq.~(\ref{eq:primitive-approx}), the momentum operator is expressed as a difference of coordinates, so that the correlation function for $\tau \ne 0$ takes the form $C_{pp}(\tau_k) = -\frac{m^2}{\hbar^2\Delta\tau^2}\langle (x_{k+1}-x_k)(x_1-x_0)\rangle$, where $x_k \equiv x(\tau_k)$, and $\tau_k = k\Delta\tau \equiv k\frac{\beta}{M}$ is proportional to the discretized imaginary time. The MC evaluation of $C_{pp}(\tau_k)$ is hampered by the fact that, when $\Delta\tau$ gets small, relative fluctuations in $(x_{i+1}-x_i)$ become large and the variance of the measured observable grows rapidly (in fact it diverges for $\Delta\tau\rightarrow0$). As the uncertainty $\delta_{MC}$ of the MC estimate of an observable $A$ is related to its variance $\sigma_A^2$ by $\delta_{MC} \propto \sigma_A/\sqrt{\tau_{sim}}$, one is therefore forced to increase the simulation time, $\tau_{sim}$, in order to achieve a given precision. 

This problem was identified early in the development of PIMC, when trying to estimate the atoms kinetic energy, which is $\propto C_{pp}(\tau=0)$. A solution was proposed in~\cite{Herman1982}: instead of directly using the above expression for  $C_{pp}(\tau_k)$, the integrals entering the correlation function can be rearranged obtaining a new estimator for $C_{pp}(\tau_k)$, with identical average but smaller variance. The new expression, known in the case of the kinetic energy as the "virial estimator", does not depend explicitly on $\Delta\tau$, and therefore does not suffer from the diverging variance associated with the "naive" estimator. 

We now show that the strategy used to obtain the virial estimator can be generalized to any correlation function involving the momentum operator~\cite{PhysRevLett.111.050406}. Specifically, we consider correlation functions of the general form involved in calculation of transport coefficients, e.~g., $C_{pF}(\tau) = \langle ( \hat{p}(\tau)\hat{F}(\tau))_s (\hat{p}(0)\hat{F}(0))_s \rangle$. Here $\hat{F}(\tau)$ is a shorthand notation for a generic local function $F(\hat{X}(\tau))$, which in the case of heat transport would be related to the potential energy. The subscript $s$ indicates that the operator product, which represents an observable quantity, is by convention made Hermitian by symmetrizing the operator, as $( \hat{p}\hat{F})_s = \frac{1}{2}(\hat{p}\hat{F}+\hat{F}\hat{p})$. 

Within the primitive approximation and following this definition one obtains,
\begin{multline}
C_{pF}(\tau_k) = - \frac{m^2}{\hbar^2\Delta\tau^2} \langle (x_{k+1} - x_{k}) F(x_{k}) (x_{1} - x_{0}) F(x_0) \rangle \\
+ \frac{m}{2\hbar\Delta\tau} \langle (x_{k+1} - x_{k}) F(x_{k}) F'(x_0)\rangle - \\
+ \frac{m}{2\hbar\Delta\tau} \langle (x_{1} - x_{0}) F(x_{0}) F'(x_{k})\rangle
- \frac{1}{4} \langle F'(x_k) F'(x_0) \rangle, 
\label{eq::pF_correlation}
\end{multline}
This expression is valid for $k\ge 1$, while the case $k=0$ must be treated separately, along similar lines. 

The MC calculation of Eq.~(\ref{eq::pF_correlation}) suffers from the same numerical problem as the momentum correlations, the variance of the leading term in $1/\Delta\tau$ diverging as $\Delta\tau$ approaches zero. In order to improve the estimator, we have generalized the procedure originally used for the kinetic energy calculations ($C_{pp}(0)$), and obtain a new estimator with reduced variance for general correlation functions. We start from the first term in Eq.~(\ref{eq::pF_correlation}), which has the strongest dependence on $\Delta \tau$, and can be expressed as,
\begin{multline}
\frac{m^2}{\hbar^2\Delta\tau^2}\langle F(x_k)(x_{k+1}-x_k)F(x_0)(x_1-x_0)\rangle =\\=\frac{m^2}{\hbar^2\Delta\tau^2 Z} \int dx_0 \int dx_1 \dots \int dx_M F(x_k) (x_{k+1}-x_k) F(x_0) (x_1-x_0)\\ \rho_0(x_1-x_0; \Delta\tau)\dots \rho_0(x_M- x_{M-1}; \Delta\tau) \exp\left[-\Delta\tau \sum _{j=0}^{M} V(x_i)\right],
\end{multline}
\mh{where $\rho_0(x-y;\Delta\tau) =  \langle x \vert e^{-\Delta\tau\hat{K}}  \vert   y \rangle  \sim \exp\{-m\frac{(x-y)^2}{2\hbar^2\Delta\tau}\}$.} We now transform the set of coordinates $\{x_0, x_i\}$ to $\{x_0, y_i\}$, such that $y_i = x_{i+1}-x_i$. The constraint $x_{M} \equiv x_0$ is accounted for by introducing a term $\delta\left(\sum_{i=0}^{M-1} y_i\right)$, leading to
\begin{multline}
\frac{m^2}{\hbar^2\Delta\tau^2}\langle F(x_k)(x_{k+1}-x_k)F(x_0)(x_1-x_0)\rangle =\\=\frac{m^2}{\hbar^2\Delta\tau^2 Z} \int dx_0 \int dy_0 \dots \int dy_{M-1} \delta\left(\sum_{i=0}^{M-1} y_i\right) F\left(\sum_{i=0}^{k-1}y_i +x_0\right) \\ y_k F(x_0)y_0 \rho_0(y_0; \Delta \tau)\dots \rho_0(y_{M-1};\Delta \tau) \exp[-\Delta\tau W],
\end{multline}
with
\begin{equation}
 W = \sum _{j=0}^{M-1} V\left(\sum_{i = 0}^j y_i + x_0\right).
\end{equation}
By using the  identity:
\begin{equation}
\frac{m}{\hbar\Delta\tau}y_k \rho_0(y_k;\Delta\tau)= -\partial _{y_k} \rho_0(y_k, \Delta\tau),
\end{equation}
we can integrate by parts for the integration over $y_k$. Our next step is based on the observation that the derivative of the $\delta$ function w.~r.~t. to $y_0$ can be distributed over all coordinates, i.e., $\partial_{y_k}\delta\left(\sum y_j\right) = \frac{1}{M} \sum_i \partial_{y_i}\delta\left(\sum y_j\right)$. A second integration by parts over each of the $y_i$ variables eventually leads to
\begin{multline}
\frac{m^2}{\hbar^2\Delta\tau^2}\langle F(x_k)(x_{k+1}-x_k)F(x_0)(x_1-x_0)\rangle
 =  \\ \left\langle\frac{m}{\hbar} F(x_k)(x_1-x_0)F(x_0)\left[\frac{1}{M}\sum_{j=1}^{M-1} j 
 V'(x_j)- 
 \sum_{j=k+1}^{M-1} V'(x_j)\right]
 \right\rangle-\\- \frac{mk}{(\hbar\Delta\tau M)}\langle F'(x_k)(x_1-x_0)F(x_0)\rangle
-\frac{m}{(\hbar\Delta\tau M)}\langle F(x_k)F(x_0)\rangle. \label{eq::virial_expression}
\end{multline}
For the special case $F(x) \equiv 1$, we can show that Eq.~(\ref{eq::virial_expression}) reduces to a virial-like formula for the momenta correlations $C_{pp}(\tau_k)=\langle x_k V'(x_0)\rangle$ (see App.~\ref{sec:appendixA}). Repeating the procedure for the terms linear in $\frac{1}{\Delta\tau}$, such as the second term in Eq.~(\ref{eq::virial_expression}), we can write the correlation in a form that apparently does not depend on $\Delta \tau$ (recall that $M\Delta \tau =\beta$ is a constant). The calculations, together with the expressions appropriate for the special case $k=0$, are sketched in App.~\ref{sec:appendixA}.

In contrast with the initial expression Eq.~(\ref{eq::pF_correlation}), all terms are now  well-defined as $\Delta\tau \rightarrow 0$. We note, however, that the number of terms involved in the first part of Eq.~(\ref{eq::virial_expression}) increases linearly with $M=\beta/\Delta \tau$, so that the gain following our manipulation is not immediately obvious. The  argument that Eq.~(\ref{eq::virial_expression}) indeed leads to a variance reduction is the following: If all the $M$ contributions to the first term were independent, its variance would scale as $\Delta \tau\times M$, where $\Delta \tau$ comes from the term $\langle \vert x_1-x_0\vert \rangle$, and the factor $M$ accounts for the $M$ contributions in the sum. As the segments in the path are correlated, even if this estimate is only approximate it still indicates that the variance remains finite even for $\Delta \tau \rightarrow 0$. We explicitly verify the variance reduction  numerically for the harmonic oscillator in the following section.

We conclude this Section by emphasizing that the above derivation to improve generic estimators that involve momentum operators is by no means limited to the harmonic oscillator, but remains valid in general, in particular for the case of interacting particles.
\mh{Also, note that the derivation of the improved estimator can be adapted beyond the use of the primitive approximation \cite{RevModPhys.67.279}. Similar refinements can be expected to work when employing improved actions,  as well as within improved sampling schemes, e.g., PIMD methods based on staging or normal modes \cite{tuckerman2010statistical}, as the variance of the estimator is entirely determined by the analytical form of the kinetic energy part of the action.}
\section{\label{sec:case1}Case study I: the single harmonic oscillator}
\subsection{Computing correlation functions}
\label{sec:computing}
We now apply the methods described above to our test cases. We start by considering the canonical example of a single quantum harmonic oscillator of frequency $\omega_0$ in one dimension, with potential energy $V=\frac{1}{2} m\omega_0^2 X^2$, and focus on the time correlation function of an operator with the structure of an energy current, e.~g., $C_{pV}(\tau) = \langle (p(\tau)V(\tau))_s (p(0)V(0))_s \rangle$. \mh{Note that, while the case of the harmonic oscillator could be considered as oversimplified, this choice of observables already leads to a non trivial structure of the correlation functions. Additional examples involving a continuous distribution of frequencies and a strongly anharmonic system are treated in Sect.~\ref{sec:case2} and App.~\ref{sec:appendixD}, respectively.} 

The PIMC approach within the primitive approximation allows us to extract the values of the imaginary time correlation function $C_{pV}(\tau_k)$, at $M$ discrete time values, $\tau_k= (k-1)\beta/M$. Two main sources of inaccuracy are associated to this procedure: a systematic error, associated with the use of the primitive approximation for the density matrix, and the statistical uncertainty due to finite sampling. In the following we show how to control these issues.

\begin{figure}[t]
\center{\includegraphics[width=1. \linewidth]{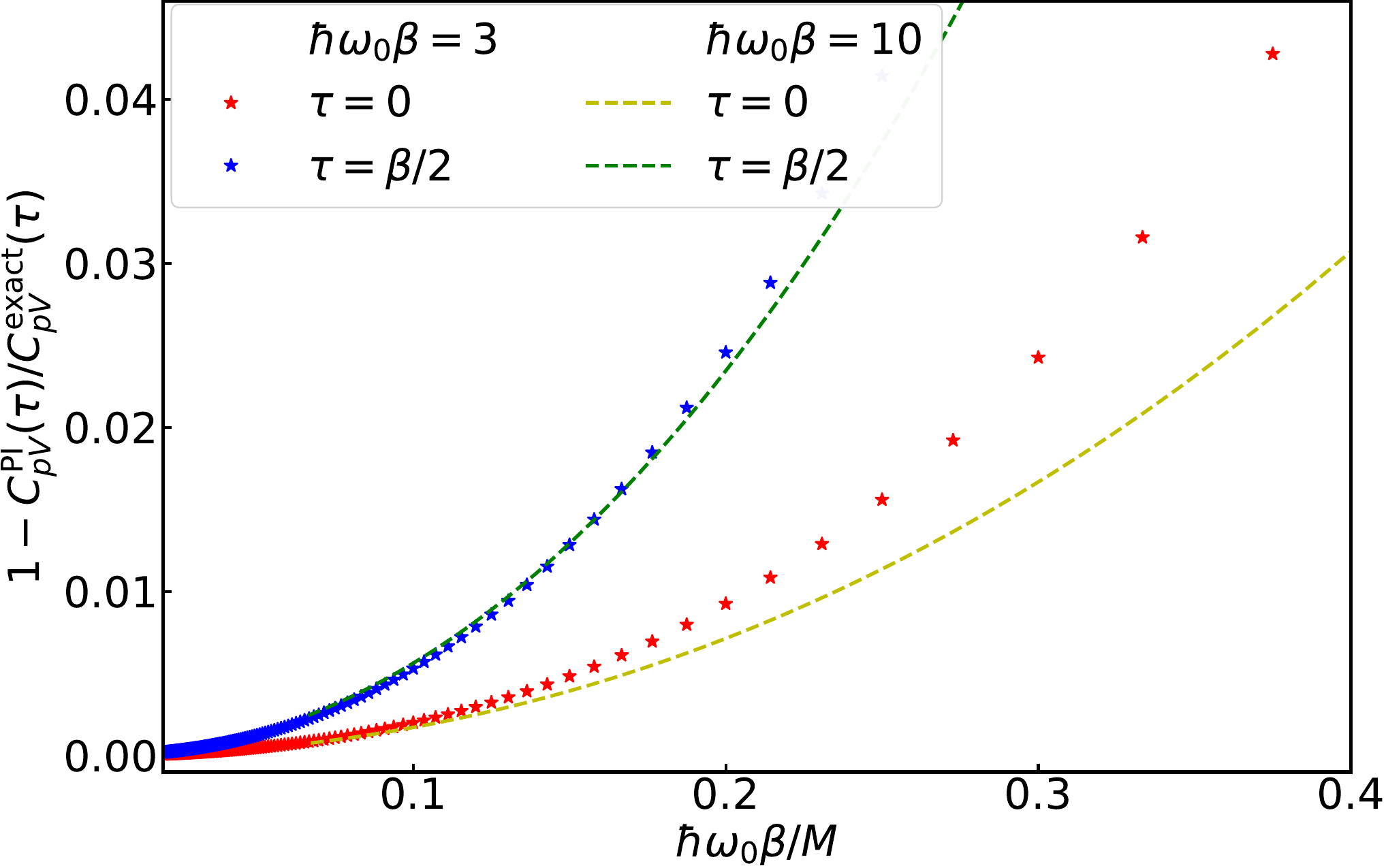}}
\caption{
Relative discretization error, $1-C_{pV}^{\mathrm{PI}}(\tau)/ C_{pV}^{\mathrm{exact}}(\tau)$, between the path integral, $C_{pV}^{\mathrm{PI}}(\tau)$,  and the exact results, $C_{pV}^{\mathrm{exact}}(\tau)$, for the energy current correlation function, as a function of $\hbar\omega_0\beta/M$. We show the data corresponding to the imaginary times $\tau=0$ and $\tau=\beta/2$, and indicate with symbols and solid lines the results for $\hbar\beta=3$ and $10$, respectively.
}
\label{fig::discretization_error_pV}
\end{figure}

For an harmonic oscillator, the systematic deviation due to the discretization of the imaginary time $\Delta\tau=\beta/M$ can be assessed directly, by comparing the result expected from the PIMC approach (which in this case can be obtained exactly) with the analytical expression for the correlation function $C_{pV}(\tau)$, which corresponds to the continuous limit $M\rightarrow \infty$. By applying the canonical formalism for the harmonic oscillator, we indeed obtain,
\begin{multline}
C^{\text{exact}}_{pV}(\tau) =\left( \frac{m\hbar^3\omega_0^3}{256}\right) \frac{1}{\sinh^3(\hbar\beta \omega_0/2)}\times \\
\left[12\cosh\left(\frac{3\hbar\beta\omega_0}{2}-3\hbar\omega_0 \tau\right)\right. \\
\left.+2\left(4e^{-\hbar\beta\omega_0}+e^{-2\hbar\beta\omega_0}+1\right)e^{\hbar\beta\omega_0} \cosh\left(\frac{\hbar\beta\omega_0}{2} -\hbar\omega_0 \tau\right) \right].
\label{eq::exact pv correlation}
\end{multline}
In order to calculate the exact expression of the correlation function within the primitive approximation of the discretized path integral, we first note that all the integrals involved in the calculation are Gaussian. By using the discretized representation for the momentum operator, one writes $C_{pV}(\tau)$ as a thermodynamic average of products of the variables $x$. Wick's theorem allows to recast such correlations $\langle x_1 \dots x_{2n}\rangle$ into products of pair correlation functions $\langle x_ix_j\rangle$ which are easily accessible as $\langle x_ix_j\rangle = A_{ij}^{-1}$. $\mathbf{A}$ is a symmetric $M \times M$ matrix, and we can write, $\langle x_ix_j\rangle =\int dX x_ix_j\text{e}^{-X^T\mathbf{A}X}$. We can therefore use numerical methods to calculate the matrix elements, as discussed in App.~\ref{sec:appendixB}. The relative difference between the two calculations is illustrated in Fig.~\ref{fig::discretization_error_pV}.
We observe that, for a sufficiently small value of $\beta/M$, the deviation is virtually not affected by a change of $\beta$.
\begin{figure}[b]
\center{\includegraphics[width=1. \linewidth]{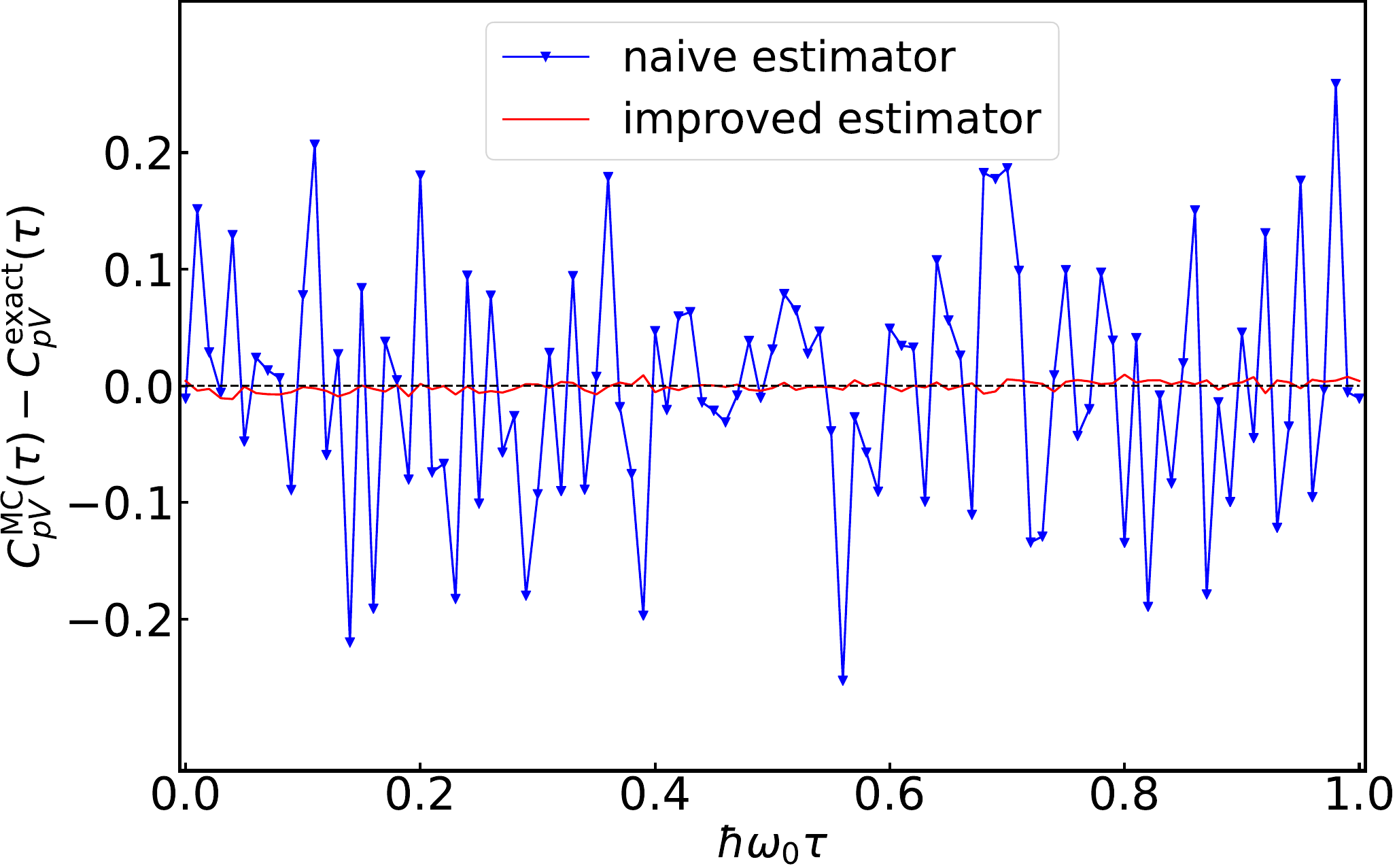}}
\caption{
Difference between the exact correlation function $C_{pV}^{\mathrm{exact}}(\tau)$ and the values obtained by Monte Carlo sampling, $C_{pV}^{\mathrm{MC}}(\tau)$, of a path with $M=100$ time slices, for $\hbar\omega_0\beta=1$, illustrating the variance reduction obtained by the improved
estimator discussed in Sect.~\ref{sec:estimators}. We show with line-points the primitive estimator and with the continuous line the improved estimator, both using the same
Monte Carlo data.
}
\label{fig::pv_correlation_beta1}
\end{figure}
\begin{figure}[t]
\center{\includegraphics[width=1. \linewidth]{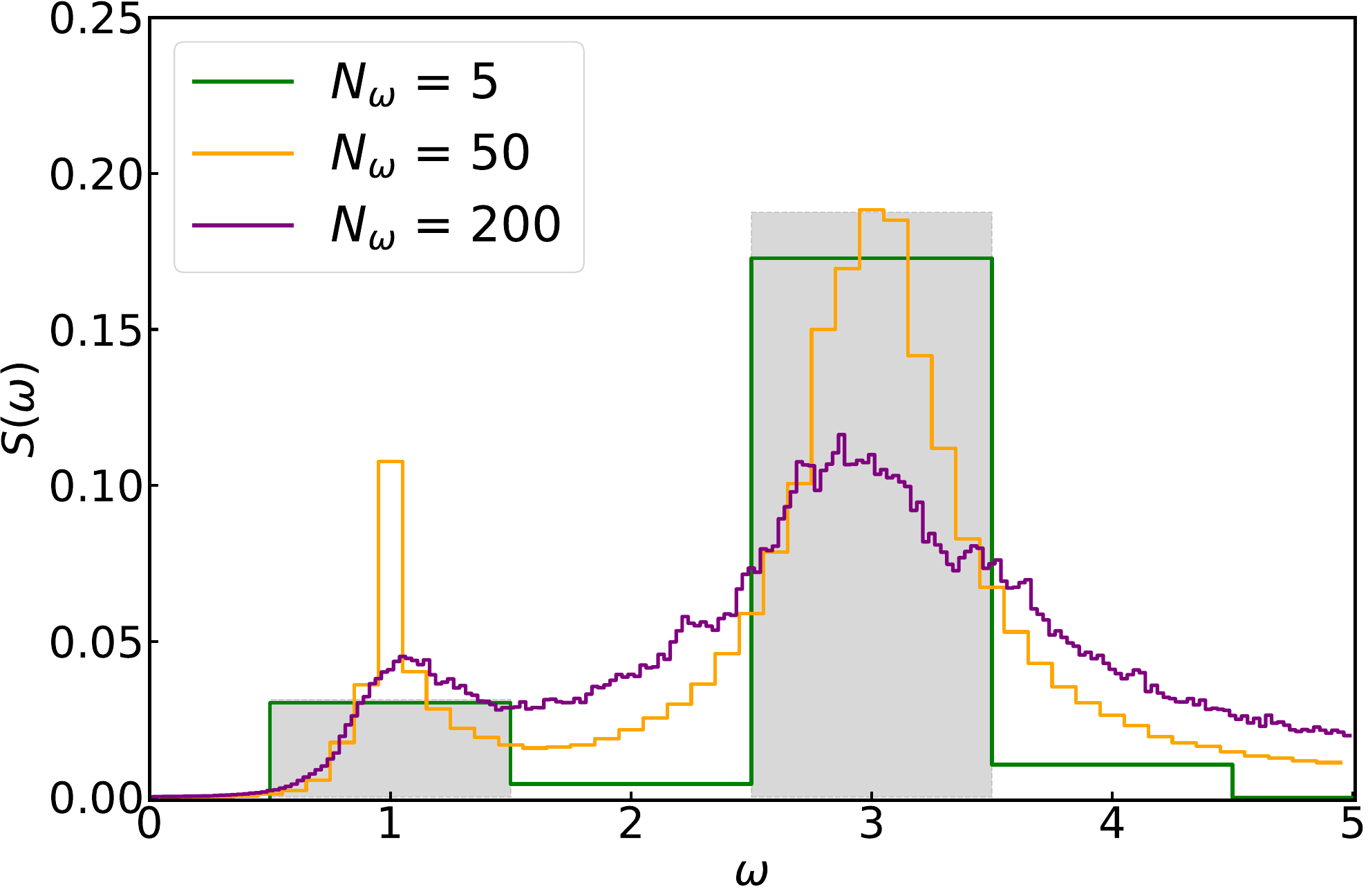}}
\caption{
Reconstruction of the spectral function associated to $C_{pV}(\tau)$ at $\beta=10$ corresponding to the indicated values for the number of delta functions in the model, $N_\omega$, and effective temperature $\Theta=1$. The area of the filled rectangles indicate the weight of the two delta-functions of the exact spectrum centered at $\omega_1=1 $ and $\omega_2=3$, corresponding to the $\Delta\omega=1$ discretization. As indicated in the text, we set $\omega_0 = 1$.
}
\label{fig::sp function b10 discretization}
\end{figure}
\begin{figure}[b]
\center{\includegraphics[width=1. \linewidth]{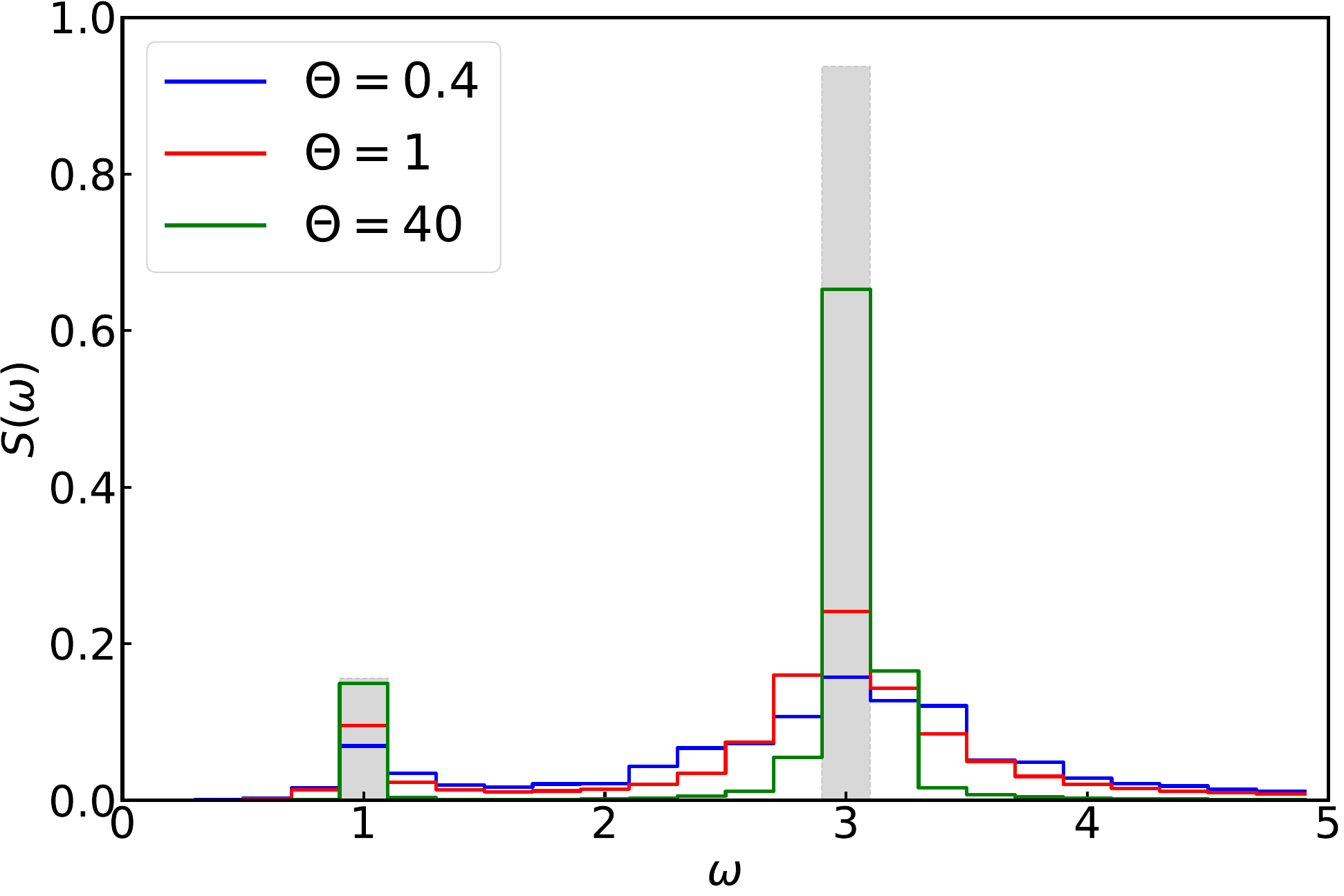}}
\caption{
Reconstructed spectra for the energy current correlation function $C_{pV}(\tau)$ at $\beta=10$, with $N_\omega=25$ and at the indicated values of $\Theta$. The filled rectangles are centered at the positions of the two delta-functions of the exact spectrum, with an area corresponding to their respective weights.} 
\label{fig::sp function b10 theta}
\end{figure}

In addition to this quantitative estimate, it is important to note that, for this system, the discretization preserves the qualitative shape of the correlation functions. One can show (see App.~\ref{sec:appendixC}) that the calculation using a finite but large $M$ corresponds to the exact result ($M\to\infty$) for slightly shifted oscillator strength and inverse temperature. The Trotter error therefore only introduces small quantitative deviations in the spectral density, but does not give rise to spurious qualitative features such as a broadening of the spectral lines. 

\mh{Also, our case studies below are performed by employing an imaginary time discretization $\tau=\beta/M=0.01\omega_0^{-1}$, a choice primarily dictated by the need to control the error associated to the primitive approximation. It also limits, however, the resolution of the imaginary time correlation function and, consequently, that of the reconstructed spectral function, especially at high frequencies. We will comment below how this potential bias can be addressed within the verification process. In  general, since the high frequency asymptotics is governed by sum rules, it is often most conveniently dealt with by computing leading terms of the short time Taylor expansion.}

We next focus on the second source of error affecting the PIMC calculation: limited sampling. Indeed, error bars corresponding to average values are obtained by estimating the variance of the observable, which decreases as $\tau_\text{sim}^{-1/2}$, with $\tau_\text{sim}$ the simulation time. For a given $\tau_\text{sim}$, the quality of the result therefore crucially depends on the variance of the estimator. We illustrate this point in Fig.\ref{fig::pv_correlation_beta1}, by comparing calculations for the energy current correlation function, $C_{pV}$, using the  naive estimator, Eq.~(\ref{eq::pF_correlation}), and the improved version of Eq.~(\ref{eq::virial_expression}). The data of Fig.~\ref{fig::pv_correlation_beta1} clearly show that the virial estimator leads to a spectacular improvement compared to the naive one, with a statistical error that is now comparable to the systematic one resulting from the discretization. 
\begin{figure}[t]
\centering
\includegraphics[width=1. \linewidth]{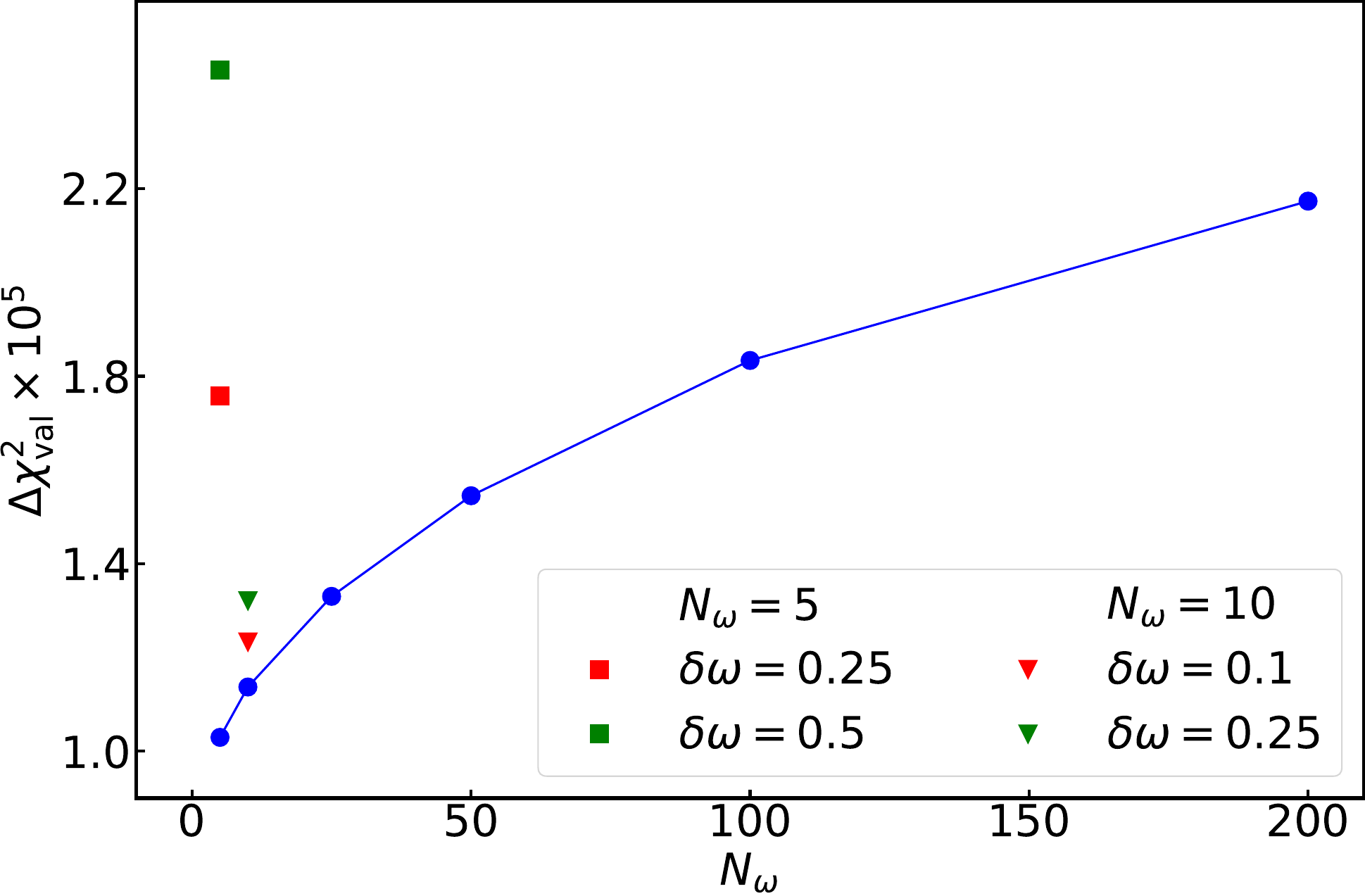}
\caption{
\fin{$N_\omega$-dependence of $\Delta\chi^2_{\mathrm{val}} \equiv \chi^2_{\mathrm{val}} - \chi^2_0$ extracted from the validation step of the reconstructed spectral functions for $C_{pV}(\tau)$, at $\beta=10$ and with $\Theta=1$. Squares and triangles correspond to shifted grids, at the indicated values of $N_\omega$ and $\delta\omega$. The value of $\chi^2_0$ corresponds to $\chi^2_{\mathrm{val}}$ for $N_\omega=5$ and  $\Theta=50$ (see Fig.~\ref{chi2 valid b10 table}, for a better comparison all values were shifted by $10^{-5}$).}
}
\label{fig::chi valid b10 discretization}
\end{figure}
\begin{figure}[b]
\center{\includegraphics[width=1. \linewidth]{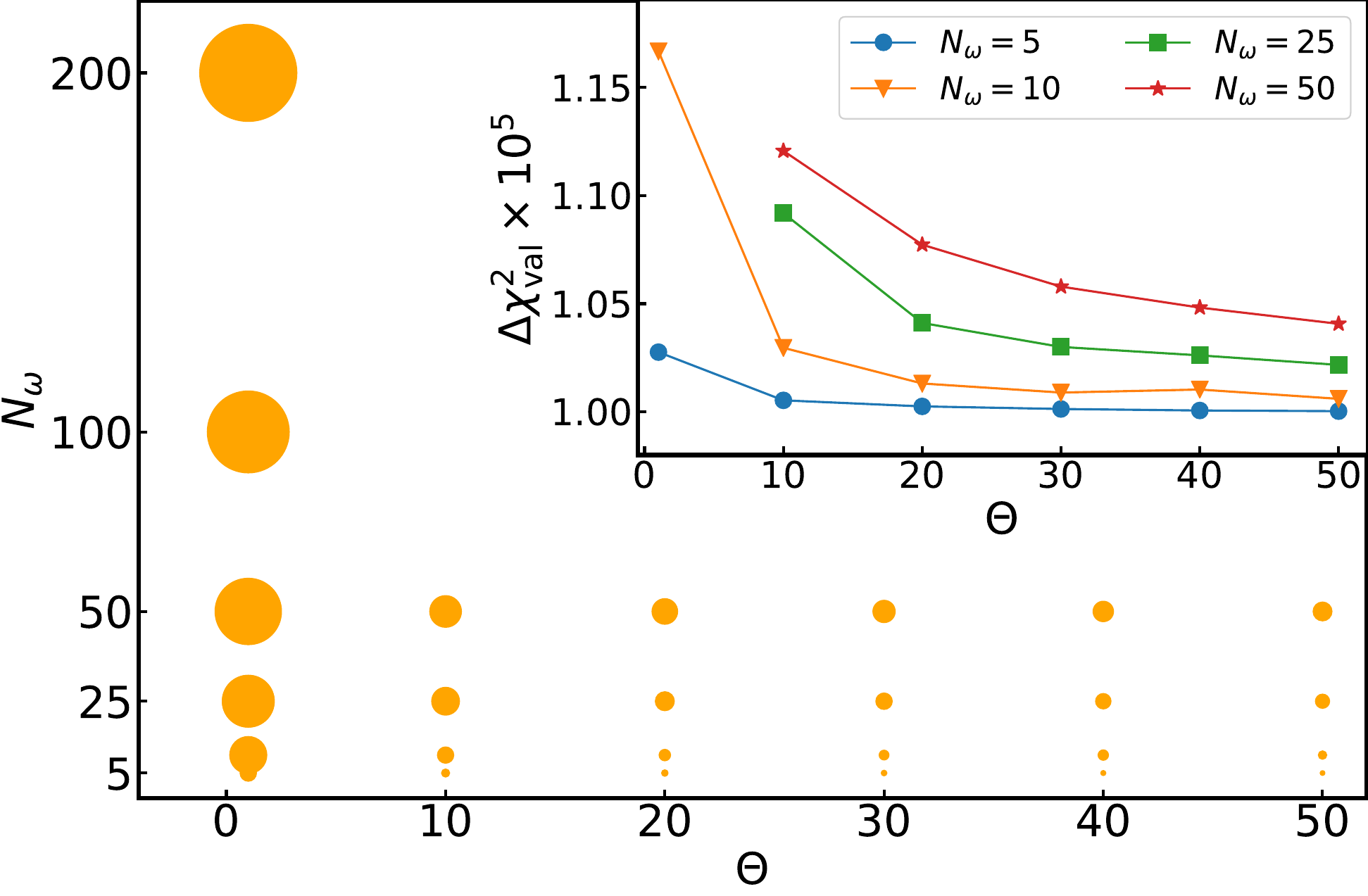}}
\caption{ 
Main panel: Comparison of the $\chi^2_{\mathrm{val}}$ obtained from our validation of the reconstructed spectral function for various values of $\Theta$ and $N_\omega$ for $C_{pV}(\tau)$ at $\beta=10$. The area of the circles is proportional to the corresponding value of $ \chi^2_{\mathrm{val}}$. Inset: \fin{ $\Delta\chi^2_{\mathrm{val}} \equiv \chi^2_{\mathrm{val}} - \chi^2_0$ as a function of the effective temperature $\Theta$, at the indicated values of $N_\omega$. The value of $\chi^2_0$ corresponds to $\chi^2_{\mathrm{val}}$ for $N_\omega=5$ and $\Theta=50$. For the purpose of a better representation, all values were shifted by $10^{-5}$. 
}
}
\label{chi2 valid b10 table}
\end{figure}
\begin{figure}[t]
\center{\includegraphics[width=1. \linewidth]{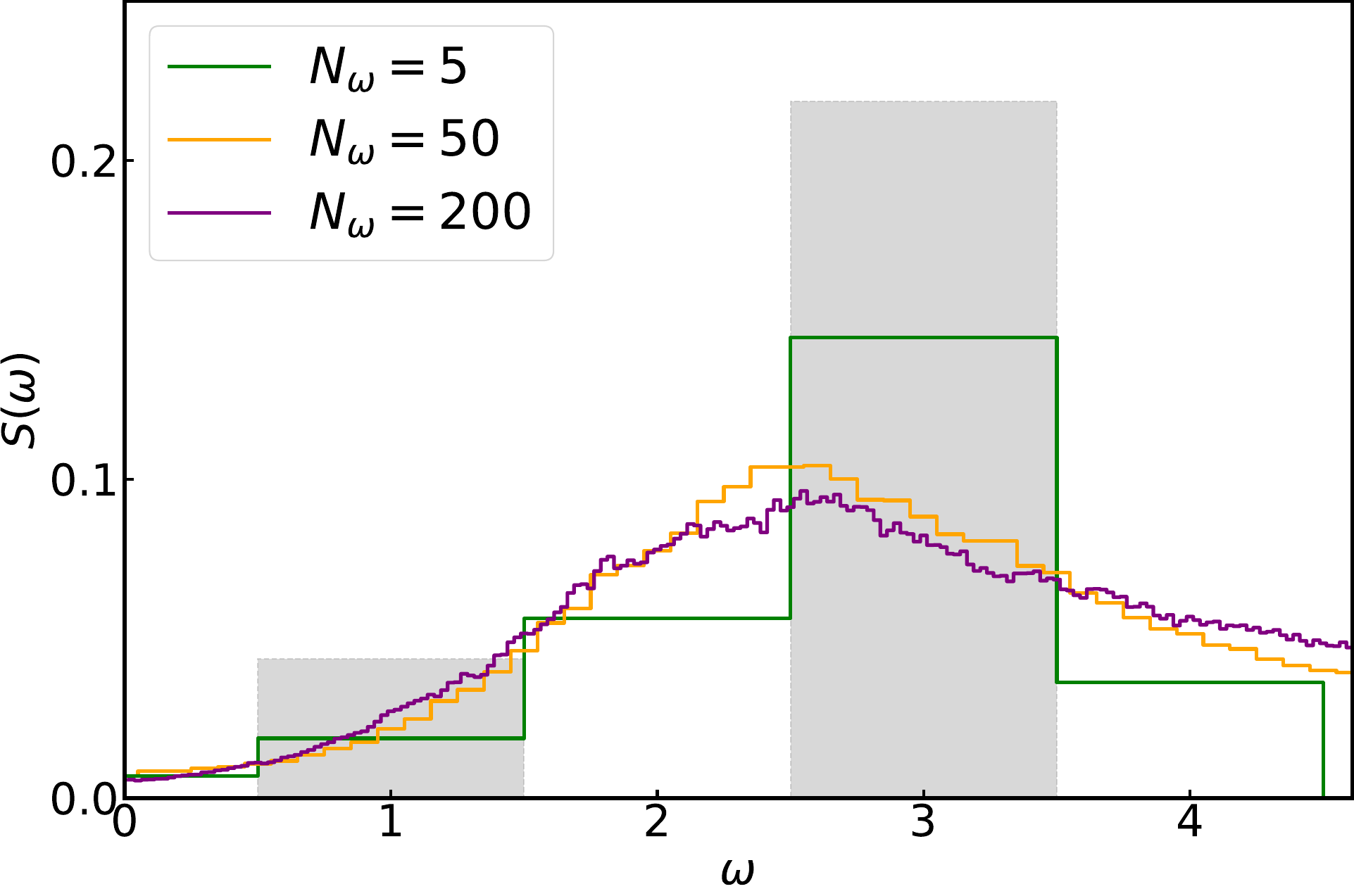}}
\caption{
Spectral reconstruction for $C_{pV}(\tau)$ at $\beta=3$, obtained at the indicated values of the discretization, $N_\omega$, for a fixed $\Theta=1$. The filled rectangles are centered at the positions of the two delta-functions of the exact spectrum for $\Delta\omega=1$, with an area corresponding to their respective weights.
}
\label{fig::sp function b3 discretization}
\end{figure}
\begin{figure}[b]
\center{\includegraphics[width=1. \linewidth]{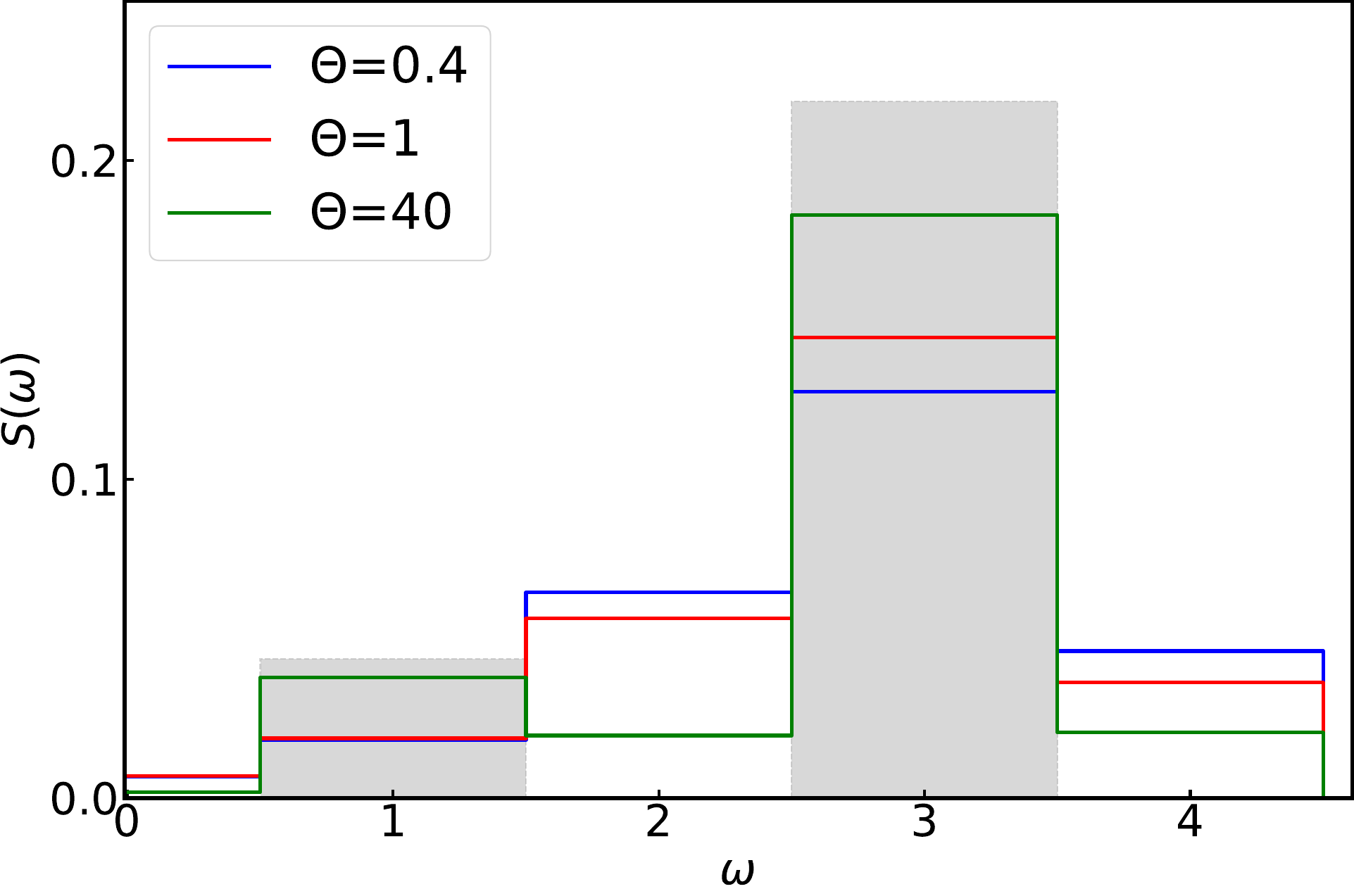}}
\caption{
Spectral reconstructions from $C_{pV}(\tau)$ at $\beta=3$ for $N_\omega=5$ using different values of $\Theta$. The filled rectangles are centered at the positions of the two delta-functions of the exact spectrum, with an area corresponding to their respective weights.
}
\label{fig::sp function b3 theta}
\end{figure}
\begin{figure}[t]	\center{\includegraphics[width=1. \linewidth]{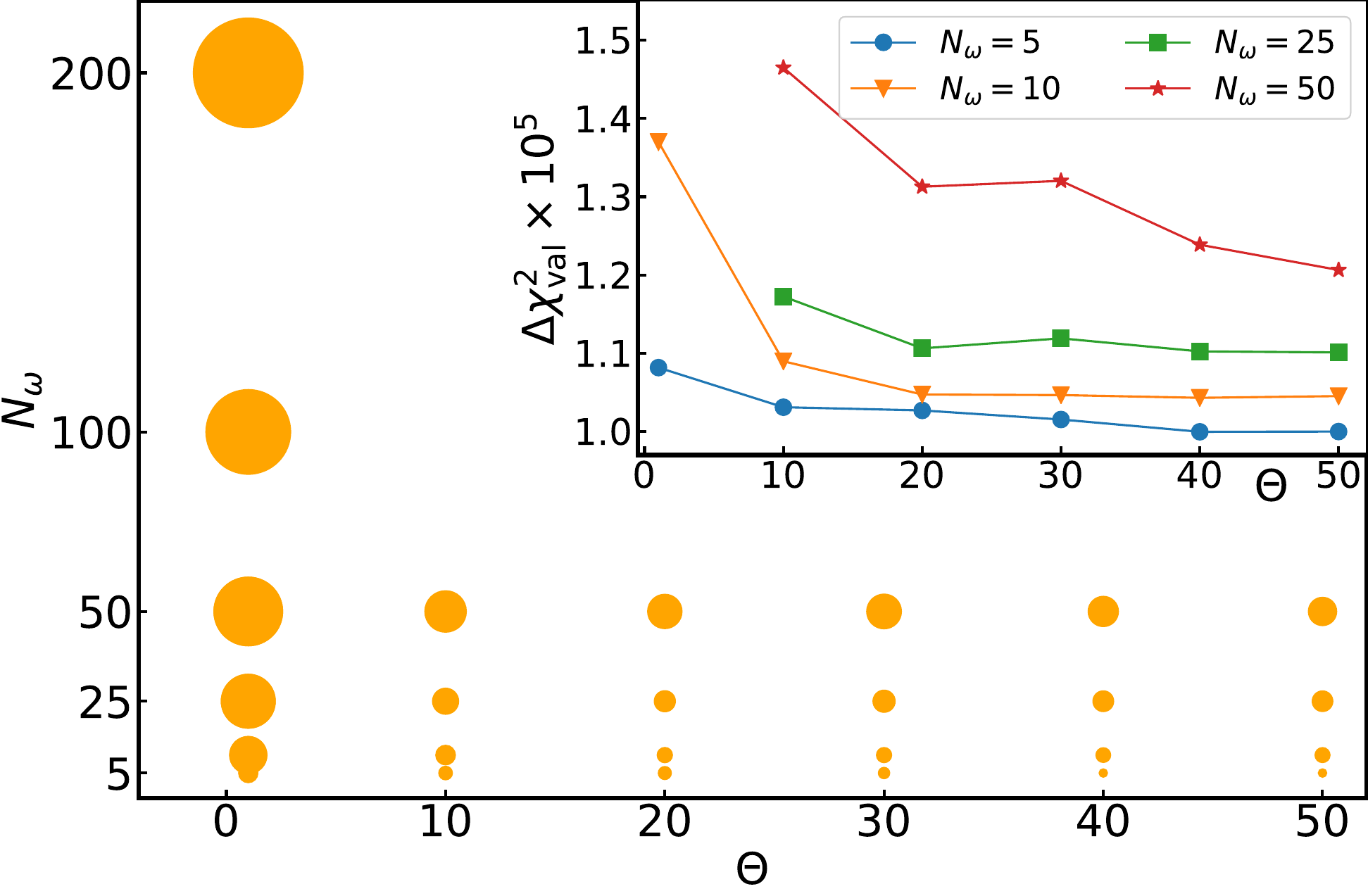}}
\caption{
Main panel: $\chi^2_{\mathrm{val}}$ from the validation procedure of the reconstructed spectral function at the corresponding values $\Theta$ and $N_\omega$ for $C_{pV}(\tau)$ at $\beta=3$. The area of the circles is proportional to the value of $ \chi^2_{\mathrm{val}}$. Inset: \fin{ $\Delta\chi^2_{\mathrm{val}} \equiv \chi^2_{\mathrm{val}} - \chi^2_0$ as a function of the effective temperature $\Theta$, at the indicated values of $N_\omega$. The value of $\chi^2_0$ corresponds to $\chi^2_{\mathrm{val}}$ for $N_\omega=5$ and $\Theta=40$. For the purpose of a better representation all values were shifted by $10^{-5}$. 
}
}
\label{chi2 valid b3 table}
\end{figure}
\subsection{The inversion problem}
\label{subsec:inversion}
We now use the reconstruction procedure outlined in Sect.~\ref{sec:inversion} to extract the frequency spectrum for the correlation functions obtained in Sect.~\ref{sec:computing}. In order to perform a reconstruction one needs both to define the set of parameters that expresses the spectral density in Eq.~(\ref{eq::correlation fit}) and in the integration measure of Eq.~(\ref{eq:sme2}), and to chose the effective inverse temperature $\Theta$. \mh{(To simplify the notation in the considered examples, we set $\omega_0=m=\hbar=1$.)} In the following, we use a discretized model of the spectral density, which is described as a sum of $N_\omega$ delta-functions in the $\omega$-space, see Eq.~(\ref{eq:sme1}). Specifically, we consider a regular grid of $\omega$-values defined on the interval $[0, 5]$, with a fixed spacing between points, $\Delta\omega=5/N_\omega$. In addition, we will consider the possibility of a global shift of the grid by $\delta\omega < \Delta \omega$. Unless specified otherwise, $\delta\omega=0$, and we fix the origin of the grid in $\omega=0$.

The exact expression for the time correlation function, Eq.~(\ref{eq::exact pv correlation}), implies that $C_{pV}(\tau)$ decays exponentially with $\tau$ in the interval $[0,\beta/2]$, with a decay rate $\mathcal{O}(1)$. Larger values of $\beta$ therefore lead to a larger amplitude in the decay, with the consequence that the contribution of different frequencies can be more easily resolved for larger $\beta$'s. In short, a correlation function of the form $[\exp(-\tau) +\exp(-3\tau)]$ will be hard to distinguish from $2\exp(-2\tau)$ if data are only available in the interval $ [0,1]$. Resolving the two frequencies $\omega_1=1$ and $\omega_2=3$ is therefore essentially impossible if $\beta/2 <1$.

In order to illustrate this point, we calculate and analyze the spectral function for the energy current correlation functions at the two inverse temperatures $\beta=3$ and $10$, with an imaginary time discretization $\Delta\tau = 0.1$. With this value of $\Delta\tau$, the systematic discretization error is smaller than the statistical error for our simulation time, so it can be safely neglected. The main constraint for the reconstruction comes from the imaginary time interval $[0, 1]$. The relative error of the MC data corresponding to these values of $\tau$ is of $\mathcal{O}(10^{-2})$. For larger $\tau$ the relative error becomes comparable with the data due to the fact that $C_{pV}(\tau)$ approaches 0 with $\tau \rightarrow\beta/2$.

We start by considering the case $\beta=10$. First, we evaluate the effect of the grid size, $N_\omega$, on the reconstruction. In Fig.~\ref{fig::sp function b10 discretization} we show the spectra obtained for various values of $N_\omega$, keeping a fixed $\Theta=1$. As  mentioned above, there is no {\em a-priori} argument guiding the most appropriate parametrization of the spectrum. In the following we analyze the accuracy of the spectral reconstruction by comparing the values of $\chi^2_{\mathrm{val}}$ defined in Eq.~(\ref{eq::chi2 validation}), using an independent test data set. This is obtained within an additional MC simulation of the correlation function, with the same parameters as the original one. We also consider a data set of the same size, $P'$, as the one that was used to produce $C_{pV}(\tau_k)$.
\begin{figure}[t]
\center{\includegraphics[width=1. \linewidth]{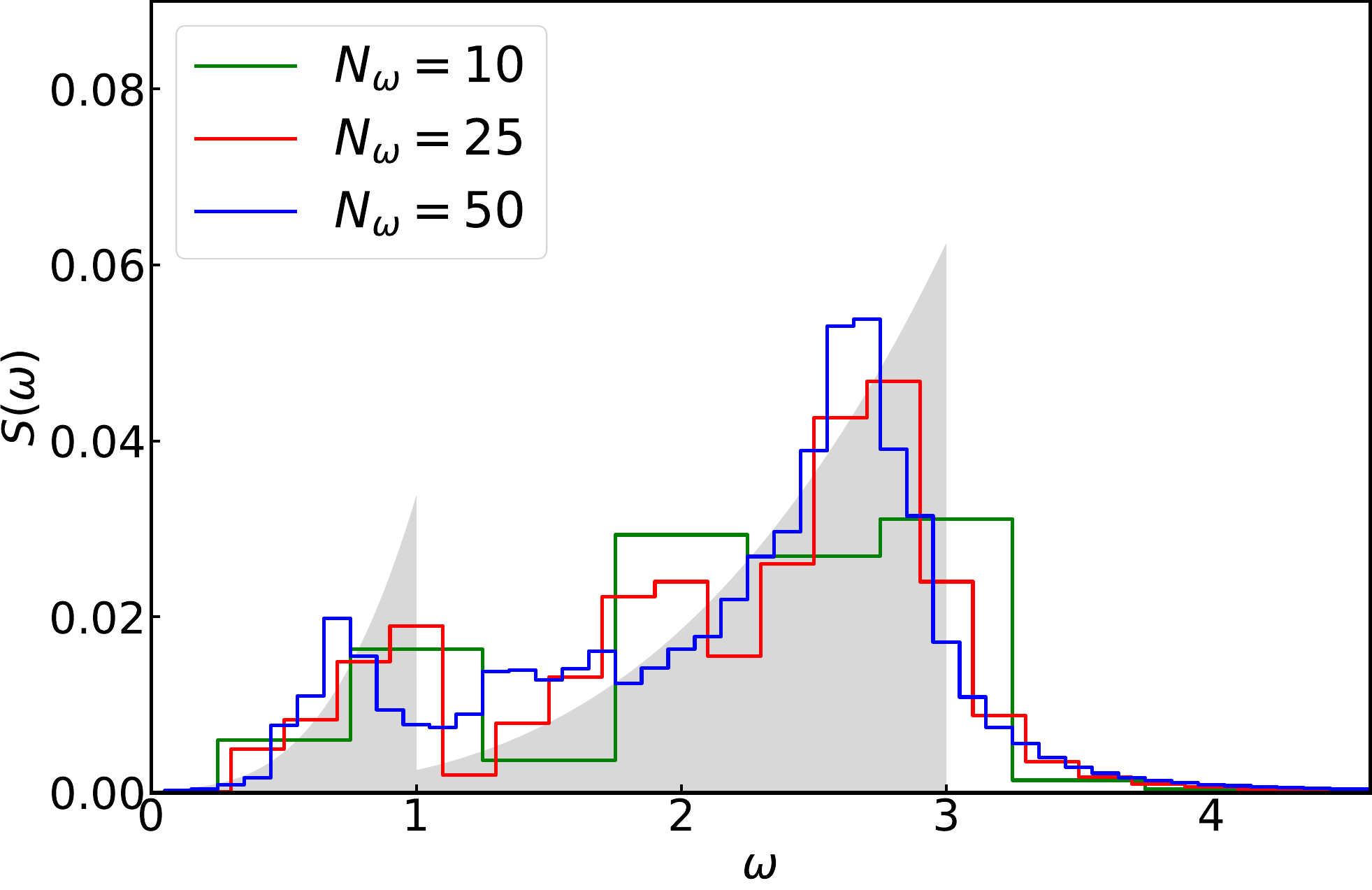}}
\caption{
Spectral reconstruction of $C_{pV}^{\text{cont}} (\tau)$ for the continuous distribution of oscillator frequencies, at the indicated values of the discretization $N_\omega$, at fixed $\Theta=1$. The shaded area indicates the exact spectral function.
}
\label{fig::contin spectrum discr}
\end{figure}
\begin{figure}[b]
\center{\includegraphics[width=1. \linewidth]{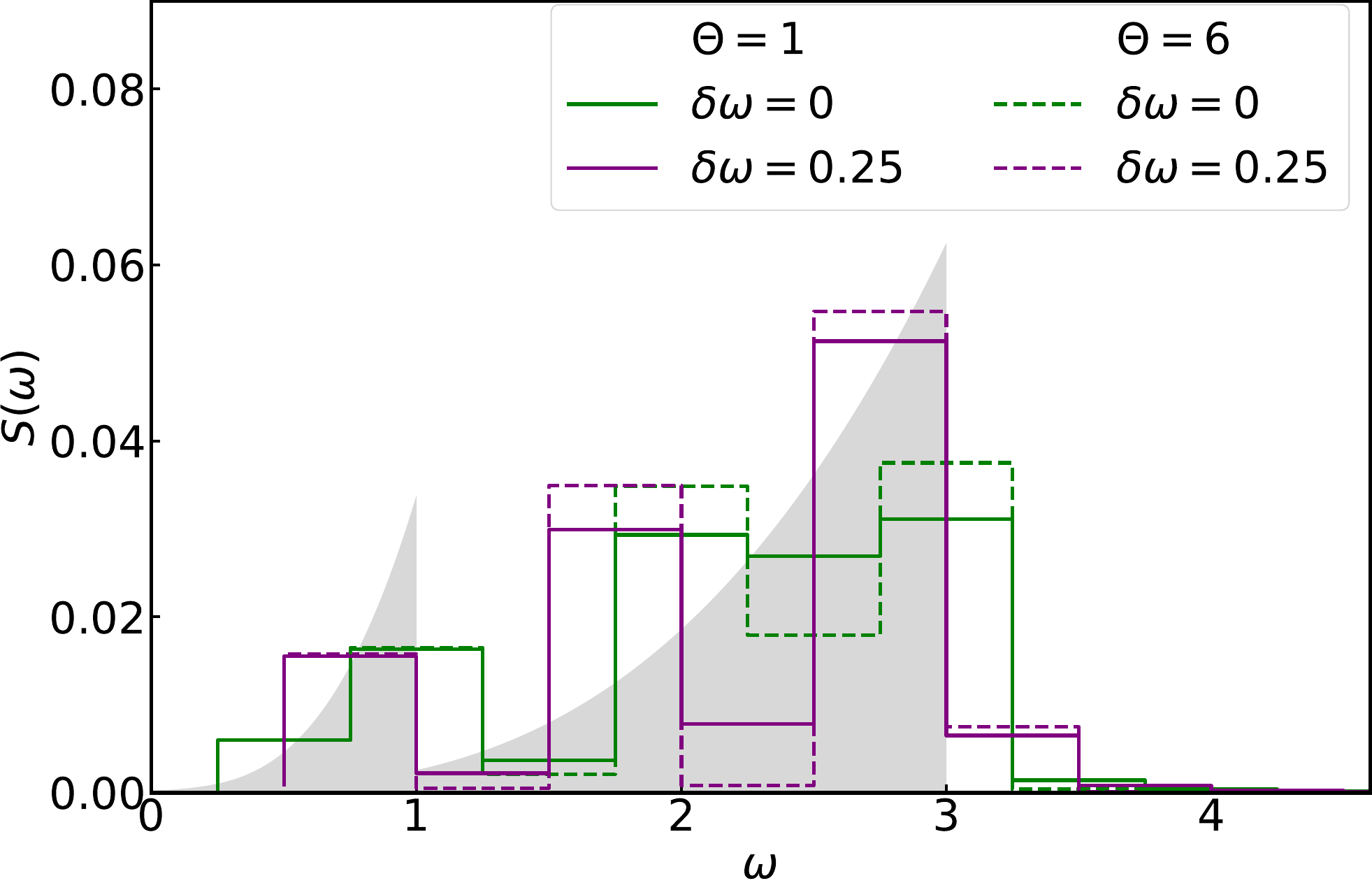}}
\caption{
Spectral reconstruction of $C_{pV}^{\text{cont}} (\tau)$ for the continuous distribution of oscillators, for $N_\omega=10$ and $\Theta=1$ and $10$, respectively. Here we compare the results pertaining to a grid shifted by $\delta\omega= 0.25$ to those with $\delta\omega=0$, the usual (not shifted) case. The shaded area indicates the exact spectral function.
}
\label{fig::contin spectrum shift theta=1 and theta=10}
\end{figure}

In Fig.~\ref{fig::chi valid b10 discretization} we show $\chi^2_{\mathrm{val}}$ as a function of the number of grid points. Clearly, increasing the number of coefficients $A(\omega_i)$ of Eq.~(\ref{eq:sme1}) does not lead to a better spectral reconstruction. In contrast, by introducing more degrees of freedom, one increases the entropy, and the spectral weight is smeared out excessively. In Fig.~\ref{fig::chi valid b10 discretization} we also show the effect on $\chi^2_{\mathrm{val}}$ of a shift $\delta \omega$. As expected, shifting the nodes away from $\omega_1=1$ and $\omega_2=3$, which are the only frequencies present in the exact spectrum determined by Eq.~(\ref{eq::exact pv correlation}), deteriorates the accuracy of the spectrum obtained through the validation step. 

The second parameter determining the quality of the statistical maximum entropy reconstruction is the effective temperature, $\Theta$. In Fig.~\ref{fig::sp function b10 theta} we show the behaviour of the spectral function for a chosen $\omega$-grid at the indicated values of $\Theta$. As expected from Eq.~(\ref{eq:sme1}), by increasing $\Theta$ the result approaches the most probable configuration that describes the correlation function $C_{pV}(\tau)$, reducing entropic effects. In Fig.~\ref{chi2 valid b10 table} we combine the above results for different pairs of parameters ($\Theta$, $N_\omega$), and plot the corresponding $\chi^2_{\mathrm{val}}$. Our validation procedure therefore strongly points to using models with a smaller number of delta functions combined with large values of $\Theta \gg 1$ for the spectral reconstruction. Based on the comparison with the exact spectrum, this choice is also clearly the one that leads to the description of the spectrum in closest agreement with the exact prediction. We conclude that the use of $\chi^2_{\mathrm{val}}$ indeed seems to provide an unbiased estimate of the quality of the reconstruction.

We now consider the spectral reconstruction for $C_{pV}(\tau)$ at $\beta=3$, again clarifying the influence of $\Theta$ and of the lattice discretization $N_\omega$. In Figs.~\ref{fig::sp function b3 discretization} and~\ref{fig::sp function b3 theta} we show selected examples of the resulting spectra. In contrast to the case $\beta=10$, we now observe in general a much stronger broadening of the peaks, which prevents us from resolving the two peak structure for $\Theta=1$, even for sparse $\omega$-grids. However, when combining sparse grids with sufficiently large $\Theta$ in the inversion, one improves towards the correct two peaks structure, as can be seen in Fig.~\ref{fig::sp function b3 theta}. The data shown in Fig.~\ref{chi2 valid b3 table} also indicate that this choice indeed corresponds to the lowest values of $\chi^2_{\mathrm{val}}$, confirming the validity of this indicator. \mh{We also note that, for large $\Theta$, the values of $\chi^2_{\mathrm{val}}$ tend to exhibit a minimum or weak oscillations, that are probably indicative of overfitting. As a consequence, considering larger values of $\Theta$ does not further improve the result.}

\mh{We conclude the above discussion by observing that in this Section the spectral reconstruction has been based on discretized imaginary time correlation functions. For convenience, the discretization usually coincides with the imaginary time step controlling the Trotter error of the path integral. In order to estimate the influence and potential bias of the discretization on the spectral reconstruction, one can perform the verification step involving subsets of $C^\prime(\tau_k)$ at no additional cost. For the test case of the double well potential discussed in App.~\ref{sec:appendixD}, we have investigated this point explicitly. In particular, we have observed that the verification is not qualitatively affected by the change of data discretization for a reasonable range of $\Delta\tau$, apart from a shift of the minimum in $\Theta$.}

\begin{figure}[t]
\center{\includegraphics[width=1. \linewidth]{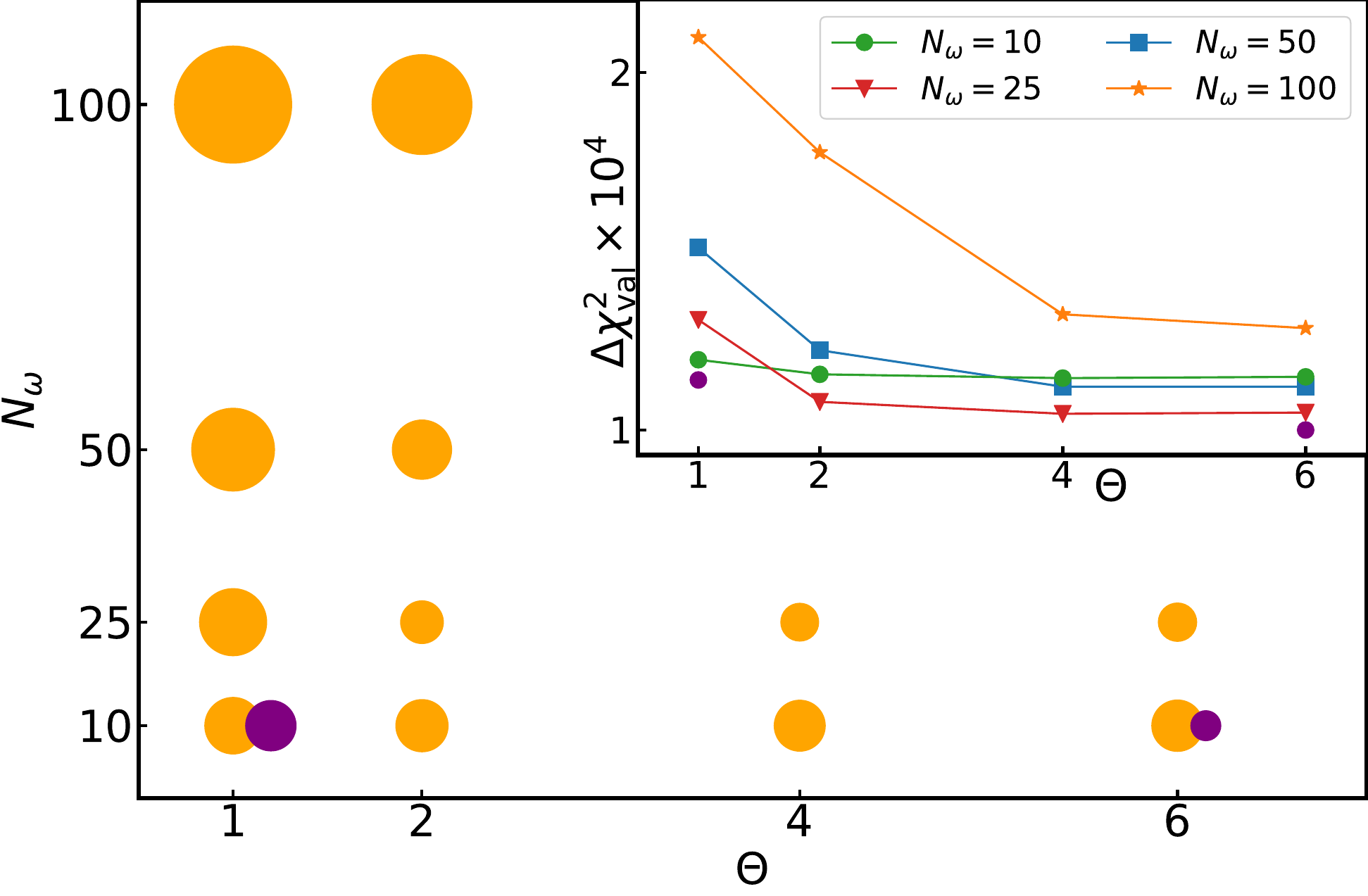}}
\caption{
Main panel: $\chi^2_{\mathrm{val}}$ from the validation procedure of spectral function at the corresponding values $\Theta$ and $N_\omega$ for $C_{pV}^{\text{cont}} (\tau)$. The area of the circles is proportional to the value of $ \chi^2_{\mathrm{val}}$. 
Purple circles correspond to the results for an $\omega$-grid shifted by $\delta\omega=\Delta\omega / 2$.
Inset: \fin{$\Delta\chi^2_{\mathrm{val}} \equiv \chi^2_{\mathrm{val}} - \chi^2_0$ as a function of the effective temperature $\Theta$, at the indicated values of $N_\omega$.
The value of $\chi^2_0$ corresponds to $\chi^2_{\mathrm{val}}$ for a shifted grid with $N_\omega=10$ and $\Theta=6$. For the purpose of a better representation all values were shifted by $10^{-4}$. }
}
\label{fig::cont spectrum valid}
\end{figure}
\section{\label{sec:case2}Case study II: continuum distribution of oscillators}
We now move to our second test model, and study the potential energy current correlation function of a system containing a large number of independent, non interacting harmonic oscillators. Considering the $C_{pV}$ of Eq.~(\ref{eq::exact pv correlation}) as a function of $\omega_0$, the correlation function for an ensemble of oscillators with a continuum of frequencies can be  written as,
\begin{equation}
C_{pV}^{\text{cont}} (\tau) = \int_0^{\omega_{cut}} d\omega_0\; C^{\text{exact}}_{pV}(\tau;\omega_0) g(\omega_0).
\label{eq::exact continous pv correlation}
\end{equation}
The form of the density of states, $g(\omega_0)$, and the value of the frequency cutoff, $\omega_{cut}$, are arbitrary. In the following we consider a Debye-like $g(\omega_0)\propto\omega_0^2$, with $\omega_{cut}=1$, and fix $\beta=10$. With this choice, the exact spectrum  for the energy current correlation is a superposition of two functions with a compact support, assuming non zero values in the range $[0,\omega_{cut}]$ and $[0,3\omega_{cut}]$, respectively. As a result, it will display two sharp discontinuities, at $\omega_{cut}$ and $3\omega_{cut}$, respectively. 
Contrary to the single oscillator case, here we do not generate the data by Monte Carlo simulation, but we rather employ the exact analytical expression, subsequently adding a Gaussian random noise with a variance proportional to the data themselves, $\sigma_k = 10^{-2}\times C_{pV}^{\text{cont}} (\tau_k)$. This variance is also used as the uncertainty to compute the $\chi^2$ of Eq.~ (\ref{eq:chi2}).

By following the same workflow discussed above for the single oscillator, we reconstruct the spectral densities for different values of $\Theta$ and number of delta functions in the model, $N_\omega$. In Fig.~\ref{fig::contin spectrum discr}, we show the influence of the discretization $N_\omega$ by fixing the canonical value $\Theta=1$. Following the same procedure as above, we calculate again $\chi^2_{\mathrm{val}}$ for the validation set by generating test correlation function from the exact result of Eq.~(\ref{eq::exact continous pv correlation}), with the same variance $\sigma_k$. The values of $\chi^2_{\mathrm{val}}$, shown in Fig.~\ref{fig::cont spectrum valid}, indicate again a more statistically sound reconstruction corresponding to sparse grids. Unfortunately, none of the curves of Fig.~\ref{fig::contin spectrum discr}, convincingly captures the sharp edges of the exact spectral density, which rather resemble two symmetrically broadened peaks. Considering shifted grids (Fig.~\ref{fig::contin spectrum shift theta=1 and theta=10}), however, as also quantitatively supported by the validation procedure, results in contrast in more asymmetric features, clearly improving the reconstruction towards the exact spectrum. Note, however, that employing sparse $\omega$-grids considerably limits frequency resolution, so that the reconstruction in the case of the continuous spectrum with its sharp discontinuities remains quite difficult.

\section{\label{sec:conclusion}Discussion and conclusions}
Here, we have examined the reconstruction of spectral functions for transport coefficients, starting from imaginary time correlation functions obtained by path integral Monte Carlo simulations. In particular, we have described a general strategy for wisely expressing improved estimators with reduced statistical variance for imaginary time correlation functions involving current or momentum operators. We have next introduced an inversion procedure based on a stochastic maximum entropy method, a Bayesian approach commonly used for such problems. The outcome of these procedures is, in general, strongly dependent on the involved parameters, as we have illustrated in the case of the harmonic oscillator spectra employing different values for the effective inverse temperature, $\Theta$, as well as different choices for the grid discretization, $N_\omega$, or offset, $\delta \omega$. Despite their apparent simplicity, the oscillator models studied here provide challenging benchmarks for the spectral reconstruction due to the sharp undamped delta-functions they contain.

Pure Bayesian approaches suggest to eliminate the parameters dependence by using a flat prior with the most general and flexible model for the spectral density, e.~g., a large value for $N_\omega$, together with $\Theta=1$ to encompass all possible solutions consistent with the data. In contrast, in our case studies we have shown that the spectra corresponding to these standard choices exceedingly suffer from the usual problems of all maximum entropy reconstructions: broadening or merging of peaks, smoothing out any sharp features in the underlying exact spectrum. 

Indeed, in practice, path integral Monte Carlo data are strongly correlated in imaginary time, undermining a true justification of the Bayesian choice $\Theta=1$. Different values of $\Theta$ may therefore be considered to efficiently approximate the true, unknown likelihood function. On the other hand, the use of flexible models for the spectral function, containing a large number of parameters, possibly introduces a large amount of entropy into the Bayesian inversion, such that different parametrizations (linear or logarithmic grids in regions where spectral densities are flat, for instance) in general strongly modify the results. The representation of a model must therefore be considered itself as a "parameter", making illusory in our view a "parameter-free" Bayesian inversion.

In this paper we have addressed exactly the above difficulties, and developed a validation procedure to quantitatively control any parameter dependence of the Bayesian inversion. Our proposal is based on the quantity $\chi^2_{\mathrm{val}}$ constructed from independent data not involved in the maximum entropy inversion, which provides an efficient and readily applicable method to select the optimal choice of parameters, corresponding to the lowest value of $\chi^2_{\mathrm{val}}$.

We have shown explicitly that the new validation step clearly identifies a discrete set of two delta functions in the case study of the single harmonic oscillator, and provides unambiguous indications towards the correct asymmetric sharp edges in the case of an underlying continuous frequency spectrum. \mh{In the case of the double well potential discussed in App.~\ref{sec:appendixD}, we demonstrate the utility of sparse frequency grids for describing discrete spectral functions. Furthermore, this example provides a possible recipe for reliable calculation of a general discrete spectrum - using sufficiently sparse grids and varying (non uniformly) the spacing between nodes to achieve an optimal reconstruction validated by the $\chi^2_{\mathrm{val}}$. } In all cases, our validation procedure eventually selects models containing just a limited number of parameters, which intrinsically limits the resolution of the reconstruction. 

Overall, combining in a consistent workflow Bayesian inversion together with an efficient validation procedure able to select model parameters and effective temperature dependence, indeed seems to offer promising perspectives for capturing qualitative and quantitative features in spectral reconstruction. \mh{We also stress that it is straightforward to integrate the proposed validation step in any existing flavor of maximum entropy reconstruction, possibly including additional prior information in the model~\cite{JARRELL1996}. In addition, the validation also provides an objective comparison of the reconstructed spectra with approximated ones, e.g., those obtained from the real-time centroid dynamics \cite{Perez2009}.}

We conclude by noting that the Green-Kubo method, combined with the harmonic theory of solids and a numerical perturbative treatment of anharmonic effects, has recently proven to be remarkably effective for the determination of heat conductivity at low temperature in systems such as amorphous silicon~\cite{Isaeva2019,Simoncelli2019}. Our hope is to extend those works to arbitrary temperatures and stronger anharmonic effects, on one hand employing path integrals to relax the assumptions underlying the perturbative treatment of anharmonicity, and on the other hand using the strategies for the spectral reconstruction developed in the present paper.
\acknowledgements
{This work has been supported by the project Heatflow (ANR-18-CE30-0019-01) funded by the french "Agence Nationale de la Recherche". Computations were performed using the Froggy platform of the
CIMENT infrastructure, which is supported by the Rhône-Alpes region (Grant No. CPER07-13 CIRA) and the project
Equip@Meso (ANR-10-EQPX-29-01) of the ANR. The Authors thank Dr. Victor Hugo Purrello for helpful technical discussions.}
\appendix
\section{Complete expression for virial-like estimators}
\label{sec:appendixA}
As we mention in Sect.~\ref{sec:estimators}, when calculating current-current correlations one needs to compute products of $p$ and $F$. As these quantities involve terms proportional to $1/\Delta \tau^2$ and $1/\Delta \tau$, their variance is quite large. In the same Section, we have demonstrated how to re-express the first one in order to obtain a more accurate MC estimator. Quantities that are linear in $1/\Delta\tau$ in Eq.~(\ref{eq::pF_correlation}) for $C_{pF}(\tau_k)$ can be re-written as,
\begin{multline}
\frac{m}{\hbar\Delta\tau}\langle F(x_k)(x_{k+1}-x_k)F^{\prime}(x_0)\rangle =\\=
\Delta\tau\left[- \sum_{i=k+1}^{M-1}\langle F(x_k)V'(x_i)F'(x_0)\rangle - \frac{k}{\Delta\tau M}\langle F'(x_k)F'(x_0)\rangle+\right.\\
\left.+\frac{1}{M} \langle F(x_k)F'(x_0)\sum_{i=1}^{M-1} iV'(x_i)\rangle\right],
\end{multline}
and,
\begin{multline}
\frac{m}{\hbar\Delta\tau}\langle F'(x_k)(x_{1}-x_0)F(x_0)\rangle =\\=
\Delta\tau\left[\frac{1}{\Delta\tau }\langle F''(x_k)F(x_0)\rangle- \sum_{i=1}^{M-1}\langle F'(x_k)V'(x_i)F(x_0)\rangle -\right. \\\left.-\frac{k}{\Delta\tau M}\langle F''(x_k)F(x_0)\rangle
+\frac{1}{M} \langle F'(x_k)F(x_0)\sum_{i=1}^{M-1} iV'(x_i)\rangle \right].
\end{multline}
When computing with Eq.~(\ref{eq::virial_expression}), we need to keep in mind that the expression is valid only for $\tau_k \neq 0$. For the case $k=0$, we can apply a similar trick finding the "virial" form,
\begin{multline}
\frac{m^2}{\hbar^2\Delta\tau^2}\langle(x_{1}-x_0)F(x_0)^2(x_{1}-x_{0}) \rangle =\\=\frac{m}{\hbar^2\Delta\tau^2 Z} \int dx_0 \int dy_0 \dots \int dy_{M-1} \delta(\sum_{i=0}^{M-1} y_i) F(x_0)^2 y_{0}y_0\\ \rho_0(y_0; \Delta\tau)\dots \rho_0(y_{M-1}; \Delta\tau) \exp(-\Delta\tau W),
\label{eq::pV0_virial_expression}
\end{multline}
where,
\begin{equation}
W = \sum _{j=0}^{M-1} V(\sum_{i = 0}^j y_i + x_0).
\end{equation}
By using the relation
\begin{equation}
\frac{m}{\hbar\Delta\tau}y_0 \rho(y_i;\Delta\tau)= -\partial _{y_0} \rho(y_0, \Delta\tau)
\end{equation}
we can re-write (\ref{eq::pV0_virial_expression}) as,
\begin{multline}
\frac{m^2}{\hbar^2\Delta\tau^2}\langle(x_{1}-x_0)F(x_0)^2(x_{1}-x_{0}) \rangle =\\= -\frac{m}{\hbar\Delta\tau Z}\int dx_0\int dy_0 \dots \int dy_{M-1} \delta(\sum_{i=0}^{M-1} y_i) \\F(x_0)^2y_0 \partial_{y_0}\rho_0(y_0; \Delta\tau)\dots \rho_0(y_{M-1};\Delta \tau)\exp(-\Delta\tau W)=\\ =-\frac{m}{\hbar}\sum_{k=0}^{M-1}\langle F(x_0)^2(x_1-x_0)V'(x_{k})\rangle + \frac{1}{\Delta\tau}\langle F(x_0)F(x_0)\rangle \\+\frac{m}{\hbar\Delta\tau Z}\int dx_0 \int dy_0 \dots \int dy_{M-1} \partial_{y_0}\delta(\sum_{i=0}^{M-1} y_i) F(x_0) \\y_0 F(x_0) \rho_0(y_0; \Delta\tau)\dots \rho_0(y_{M-1}; \Delta\tau) \exp(-\Delta\tau W). 
\end{multline}
By substituting, 
\begin{equation}
\partial_{y_0}\delta(\sum_{i=0}^{M-1} y_i) =\frac{1}{M}\sum_{j=0}^{M-1}\partial_{y_j}\delta(\sum_{i=0}^{M-1} y_i),
\end{equation}
we finally obtain,
\begin{multline}
\frac{m^2}{\hbar^2\Delta\tau^2}\langle(x_{1}-x_0)F(x_0)^2(x_{1}-x_{0}) \rangle =\\=
(\frac{m}{\hbar\Delta\tau}-\frac{m}{\hbar\Delta\tau M})\langle F(x_0)F(x_0)\rangle -\frac{m}{\hbar}\sum_{k=0}^{M-1}\langle F(x_0)^2(x_1-x_0)V'(x_{k})\rangle\\
+ \frac{m}{\hbar M} \langle F(x_0)(x_1-x_0)F(x_0)\left(\sum_{j=1}^{M-1} j V'(x_j)+ M V'(x_0)\right)\rangle. 
\end{multline}
This expression (a similar argument applies to Eq.~(\ref{eq::virial_expression})) reduces to the usual virial formula for the kinetic energy when $F=1$. Indeed, using cyclic invariance along the path, one obtains (setting for simplicity $\hbar$ and $m$ to unity):
\begin{multline}
\frac{1}{\Delta\tau^2}\langle(x_{1}-x_0)(x_{1}-x_{0}) 
\rangle =
\\ =\frac{1}{\Delta\tau}-\frac{1}{\Delta\tau M}+\frac{1}{M} \left\langle (\frac{1}{M} \sum_{j=0}^{M-1} x_j) (\sum_{j=0}^{M-1} V'(x_j))\right\rangle  - \langle x_0 V'(x_0) \rangle,
\end{multline}
By employing the classical virial theorem for the center of mass of the path, the second and third terms  cancel mutually, and we are left with the usual result for the  kinetic energy estimator.
\section{Exact computation of correlation for the harmonic oscillator with a discretized path integral}
\label{sec:appendixB}
Path integral calculations are usually excessively involved to evaluate them analytically, even if the naive discretized version of the density matrix is used. For the case of a single harmonic oscillator, however, one can find the result explicitly. Within this approximation, the partition function, $Z_{h.o.}$ of the harmonic oscillator is written as a Gaussian integral,
\begin{equation}
Z_{h.o.} = \int dX e^{-\frac{1}{2}X^T\mathbf{A} X},
\end{equation}
where $X^T$ is a short-hand form for the vector $\{x_0, x_1, ..., x_{M-1} \}$, and we have introduced the matrix, 
\begin{equation}
\mathbf{A} = \left(
\begin{array}{cccccc}
a & b & 0 & \ldots &b \\
b & a & b &\ldots &0\\
0 & b & a & \ldots &0\\
\vdots & \vdots & \ddots & \ddots & \vdots
\end{array} \right) \label{matrix A}
\end{equation}
with $a = \frac{2m}{\hbar^2\tau} + m\omega_0^2 \tau$, and $b = -\frac{m}{\hbar^2 \tau}$. For a Gaussian weight, the correlation function $\langle x(t)x(0) \rangle \equiv \langle x_ix_0 \rangle$ can obtained as
\begin{equation}
\langle x_ix_0 \rangle = (\textbf{A}^{-1})_{i0}, 
\label{matr_inv}
\end{equation}
where $\textbf{A}^{-1}$ is the inverse matrix of $\textbf{A}$. The formal expression for the partition function is therefore, 
\begin{equation*}
Z = (2 \pi)^{\frac{M}{2}} [\text{det}\textbf{A}]^{-1/2}. 
\end{equation*}

In general, the inversion of $A$ is handled by numerical methods. The diagonal terms, however, can be obtained analytically in a straightforward manner. Indeed, we can first calculates the eigenvalues $\{\lambda\}$ of the matrix $\mathbf{A}$ by posing,
\begin{equation*}
\lambda x_j = ax_j + bx_{j-1} + bx_{j+1}.
\end{equation*}
By looking for solutions of the form,
\begin{equation*}
x^{(k)}_j = e^{2 i \pi \frac{kj}{M}}
\end{equation*}
with $k\in [0, \ldots M-1]$, the corresponding eigenvalues are,
\begin{equation*}
\lambda^{(k)} = a + 2b \cos \left( \frac{2 \pi k}{M} \right ),
\end{equation*}
and the determinant is  $\det A = \prod_k \lambda^{(k)}$. 
The terms $\langle x_i^2\rangle$ are next obtained directly from
\begin{multline}
\langle x_i^2 \rangle =  \frac{1}{M}\langle \sum_i x_i^2 \rangle = \frac{1}{M} \frac{\partial}{\partial a} \log (\det A)  =\\= \sum_{k=0}^{M-1} \frac{1}{m\omega_0^2\beta + 4m \frac{M^2}{\hbar^2\beta} \sin^2 \left( \frac{\pi k}{M} \right )} =\\= \sum_{k=-M/2}^{M/2} \frac{1}{m\omega_0^2\beta + 4m \frac{M^2}{\hbar^2\beta} \sin^2 \left( \frac{\pi k}{M} \right )}.
\end{multline}
\section{Influence of the primitive approximation for the density matrix on time correlations for the harmonic oscillator}
\label{sec:appendixC}
\begin{figure}[t]
\center{\includegraphics[width=1. \linewidth]{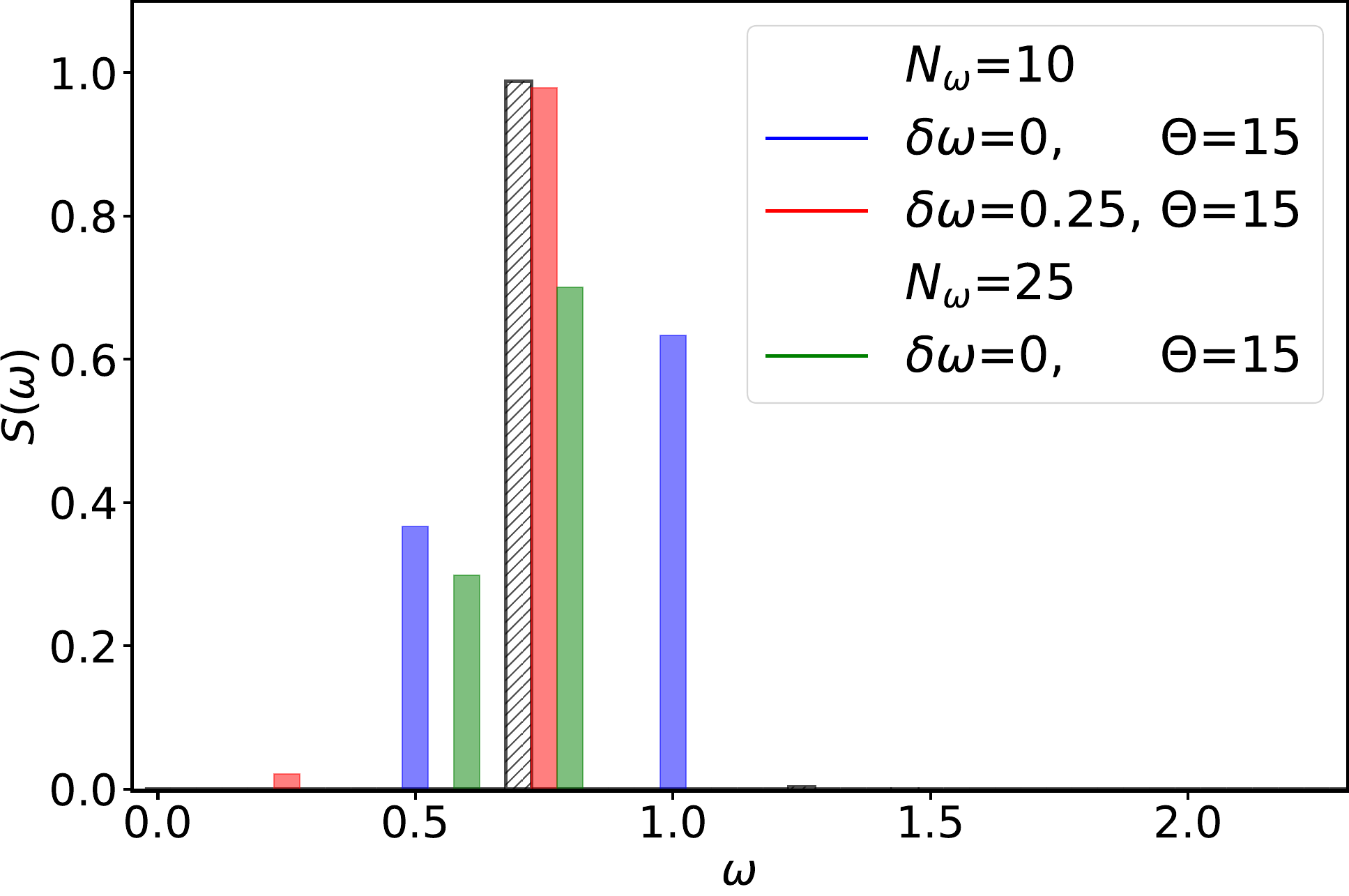}}
\caption{
Spectral reconstruction of $C^{\beta=8}_{xx} (\tau)$ for the double well potential, at the indicated values of the discretization $N_\omega$ and $\Theta$. Value of $\Theta$ corresponds to the minimum of $\chi^2_{\mathrm{val}}$ for $N_\omega=25$ model as shown in the Fig.~\ref{fig::dwell chi valid b8}. Red spectral function corresponds to a $N_\omega=10$ lattice shifted by $\delta\omega=0.25$. The exact spectrum of $C^{\beta=8}_{xx} (\tau)$ at $\beta=8$ is plotted as gray hatched area. For the purpose of illustration all delta-functions are shown with the frequency resolution $\Delta\omega = 0.01$.
}
\label{fig::dwell spectrum beta=8}
\end{figure}

\begin{figure}[t]
\centering
\includegraphics[width=1. \linewidth]{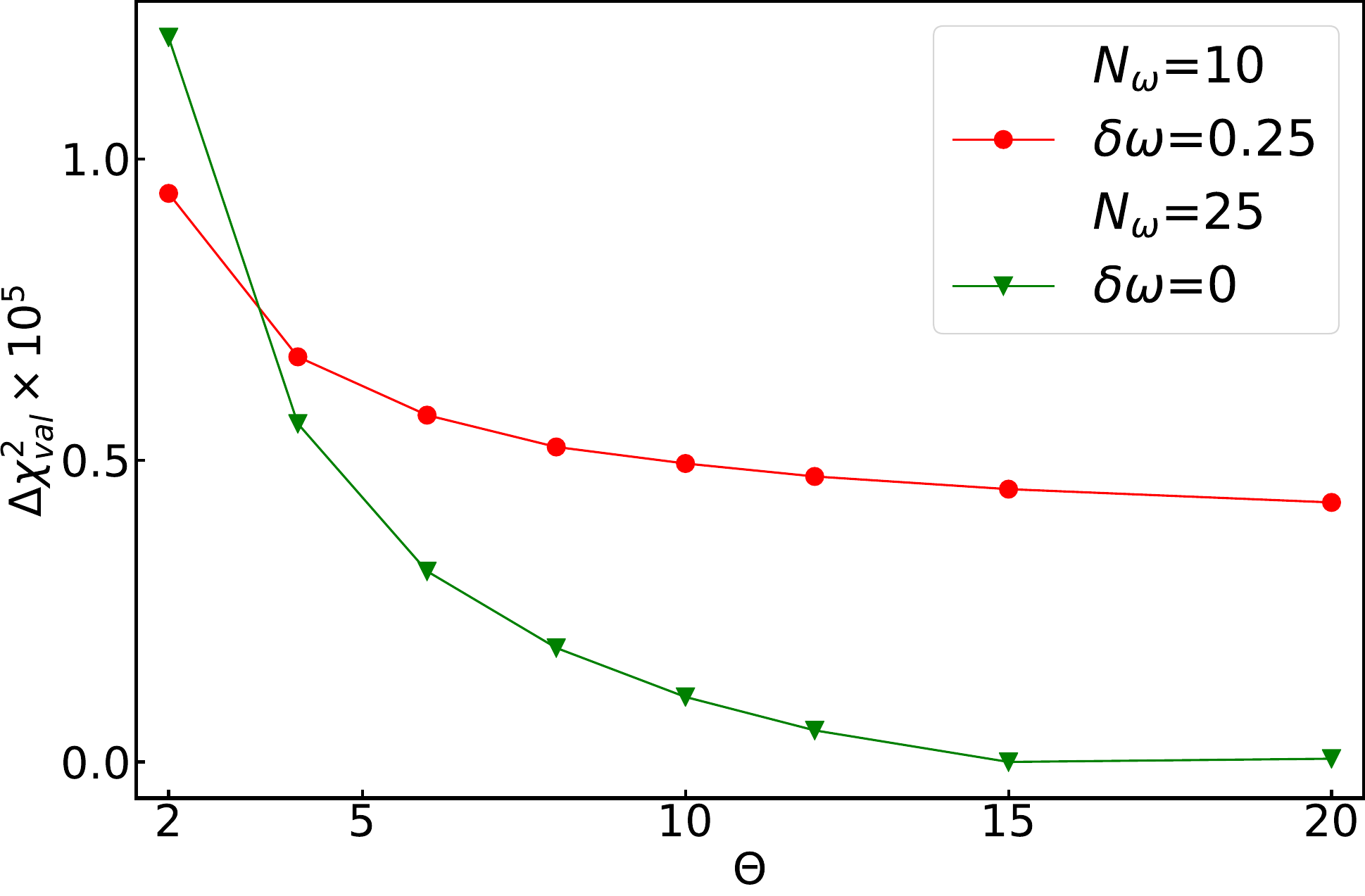}
\caption{
\fin{$\Theta$-dependence of $\Delta\chi^2_{\mathrm{val}}\equiv \chi^2_{\mathrm{val}} - \chi^2_0$ calculated from the validation step of the reconstructed spectral functions for $C^{\beta=8}_{xx}(\tau)$ for the frequency discretizations $N_\omega=10$ and $N_\omega=25$. The value of $\chi^2_0$ is $\chi^2_{\mathrm{val}}$ corresponding to $N_\omega=25$ and $\Theta=15$. Spectral functions for the non-shifted $N_\omega=10$ grid (shown at $\Theta=15$ as the blue spectrum in Fig.~\ref{fig::dwell spectrum beta=8}) provides considerably worse result and the corresponding $\chi^2_{\mathrm{val}}$ are not shown here.}
}\label{fig::dwell chi valid b8}
\end{figure}

\begin{figure}[t]
\center{\includegraphics[width=1. \linewidth]{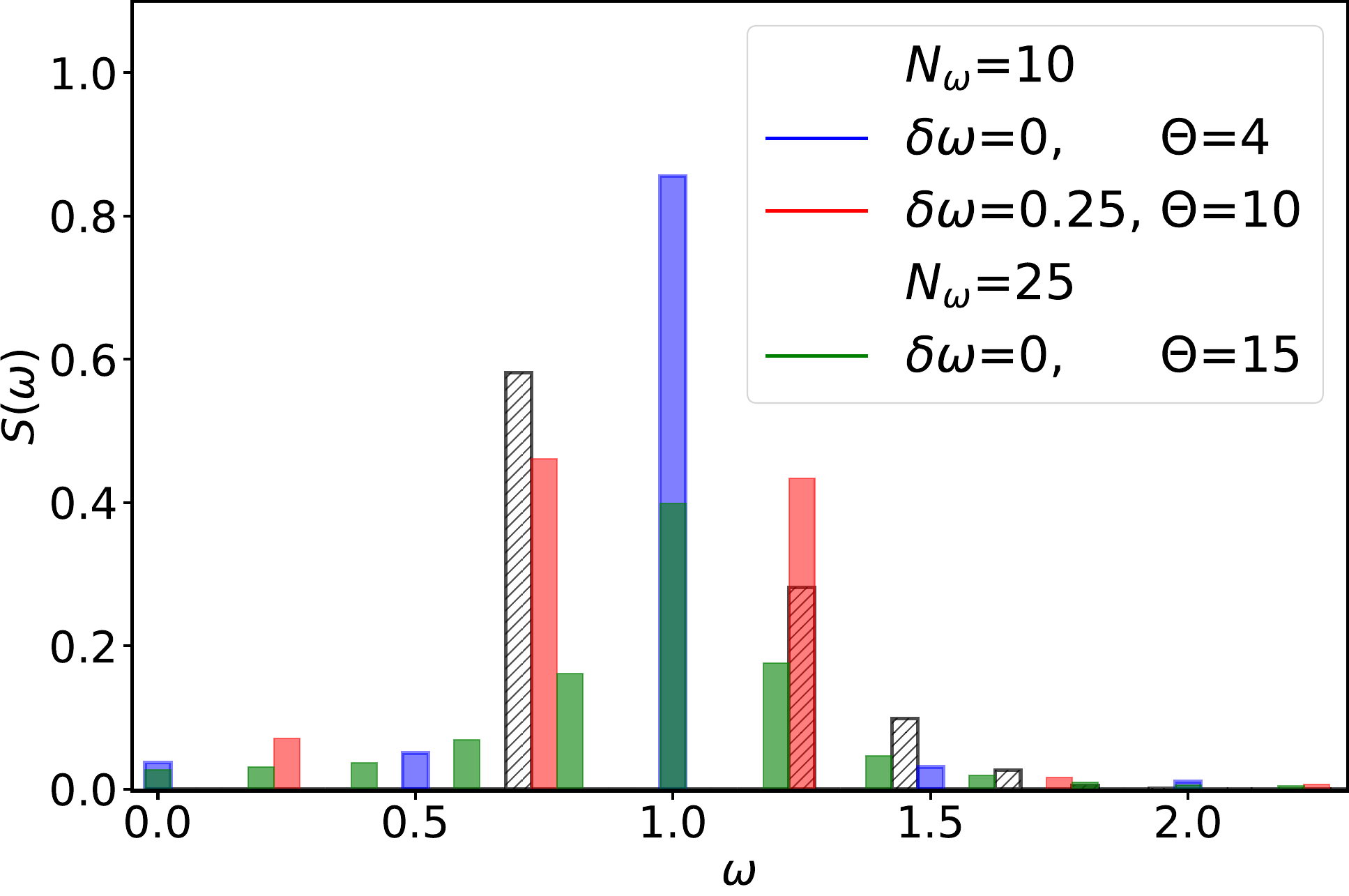}}
\caption{
Spectral reconstruction of $C^{\beta=1}_{xx} (\tau)$ for the double well potential, at the indicated $\Theta$. Values of $\Theta$ correspond to the minima of $\chi^2_{\mathrm{val}}$ for each model respectively as shown in the Fig.~\ref{fig::dwell chi valid b1}. We compare models with frequency discretization $N_\omega=10$ and $N_\omega=25$, red spectral function corresponds to a $N_\omega=10$ lattice shifted by $\delta\omega=0.25$. The gray hatched area indicates the exact spectral function at $\beta=1$. All delta-functions are shown with the frequency resolution $\Delta\omega = 0.01$.
}
\label{fig::dwell spectrum beta=1}
\end{figure}

\begin{figure}[t]
\centering
\includegraphics[width=1. \linewidth]{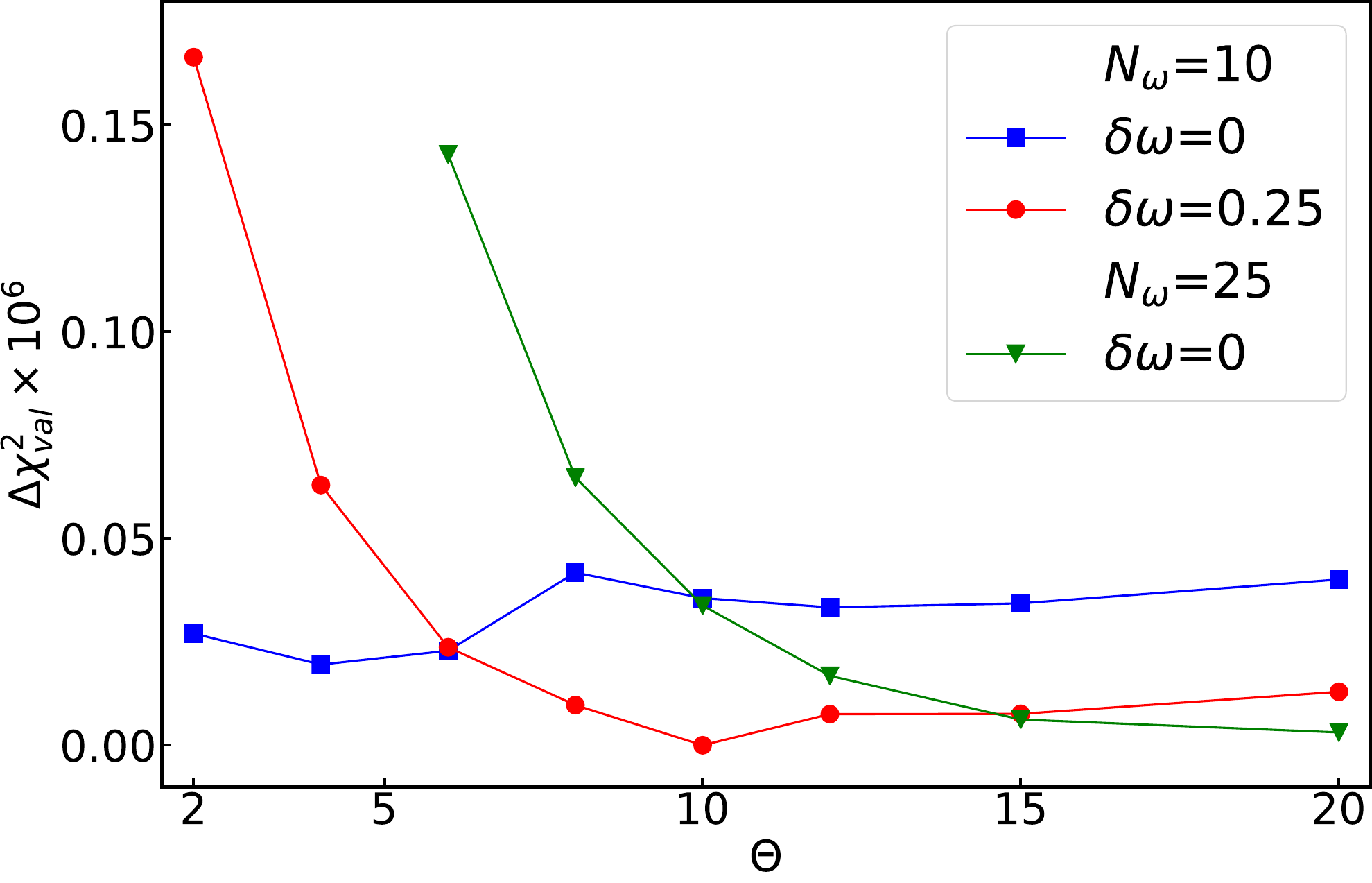}
\caption{
\fin{$\Theta$-dependence of $\Delta\chi^2_{\mathrm{val}} \equiv \chi^2_{\mathrm{val}} - \chi^2_0$ calculated from the validation step of the reconstructed spectral functions for $C^{\beta=1}_{xx}(\tau)$. The value of $\chi^2_0$ corresponds to a reconstruction with $N_\omega=10$ and a grid shifted by $\delta\omega=0.25$, at $\Theta=10$. Here we compare shifted ($\delta\omega=0.25$) and non-shifted lattices with the indicated values of $N_\omega$.}
}\label{fig::dwell chi valid b1}
\end{figure}

In the case of the harmonic oscillator, the density matrix,  $\rho(X,Y,\tau) = \langle X \vert e^{-\tau\hat{H}}  \vert   Y \rangle $ can be computed exactly, obtaining
\begin{multline}
\rho(X,Y,\tau)
= \sqrt{\frac{m\omega}{2 \pi \hbar \sinh (\hbar\omega\tau)}}
\exp
\left\{-\frac{m\omega}{4\hbar} f(X,Y)\right\},
\end{multline}
with
\begin{equation}
f(X,Y)=  (X+Y)^2 \tanh\left(\frac{\hbar\omega\tau}{2}\right) 
+ (X-Y)^2 \coth \left(\frac{\hbar\omega\tau}{2} \right).
\end{equation}
Comparing this expression with Eq.~(\ref{eq:primitive-approx}) specialized to the harmonic oscillator, we realize that the functional dependence on $X$ and $Y$ is identical for both the exact and the approximate expressions. As a consequence, a path integral simulation of an oscillator of frequency $\omega$  using the primitive approximation with an imaginary time step $\tau$, will sample the same configurations as an exact calculation with a frequency $\omega'$ and a step $\tau'$, provided that we meet the conditions, 
\begin{equation}
 \frac{\hbar \omega'}{2} \coth\left( \frac{\hbar\omega'\tau'}{2}\right) = \frac{1}{\tau} (1   + \frac{\hbar^2\omega^2\tau^2}{4} ),
\end{equation}
and, 
\begin{equation}
 \hbar\omega' \tanh\left(\frac{\hbar\omega'\tau'}{2}\right)=  \frac{(\hbar\omega)^2\tau}{2}.
\end{equation}
For a given value of $\hbar\omega\tau$ and $\hbar\omega$, these equations have a unique solution for $\omega'$ and $\hbar\omega'\tau'$. As a consequence, a calculation only involving the configurations sampled by the path (and not the normalisation of the density matrix) will correspond to the exact result for the shifted frequency $\omega'$, and inverse temperature $\beta'=M\tau'$, with a relative shift of $\mathcal{O}(\hbar\omega\tau)^2$. 
\begin{table}
 \begin{tabularx}{0.4\textwidth} { 
  | >{\centering\arraybackslash}X 
  | >{\centering\arraybackslash}X | }
 \hline\hline
 $\omega_1$ & 0.72478  \\
 \hline
 $\omega_2$ & 1.25495  \\ 
 \hline
 $\omega_3$ & 1.46154 \\
 \hline
 $\omega_4$ & 1.65143 \\
 \hline
 $\omega_5$ & 1.80533 \\
 \hline\hline
\end{tabularx}

\caption{Lowest excited states frequency values, $\omega_n = E_n - E_{n-1}$, of the spectrum of $C^{\beta}_{xx} (\tau)$ for the double well potential.}
 \label{table::dwell frequencies}
\end{table}
{\color{black} \section{Spectral function calculation for a double well potential}
\label{sec:appendixD}
In order to illustrate that our discussion is not intrinsically limited to harmonic oscillators, we examine an additional classical benchmark example~\cite{Perez2009} with strongly anharmonic features. Let us consider the Hamiltonian describing a particle trapped in a double well potential,
\begin{equation}
\hat{H} = \frac{\hat{p}^2}{2m} - \alpha \hat{x}^2 + \frac{1}{2}\gamma \hat{x}^4, \label{eq::dwell_ham}
\end{equation}
and we choose $\alpha=1$ and $\gamma=1$. We can obtain the energy spectrum and the correlation functions of interest by numerical diagonalisation of $\hat{H}$. In the following we consider the position correlation function, $C_{xx}^{\beta}(\tau) = \langle \hat{x}(\tau)\hat{x}(0)\rangle$.
The corresponding spectral function can be expressed analytically as,
\begin{equation}
S(\omega) = \frac{1}{Z}\sum_{n, m} \text{e}^{-\beta E_n} |\langle n| \hat{x}|m\rangle|^2 \delta(\omega - E_m + E_n).
\label{somega}
\end{equation}

Starting from the exact correlation function, we generate an extended data set $\tilde{C}(\tau_k)$ with an artificial Gaussian noise of variance $\sigma_k = 10^{-3}\times \tilde{C}_{xx}^{\beta} (\tau_k)$. We next employ the calculation scheme detailed in Sect.~\ref{sec:inversion}, and apply the $\chi^2_{\mathrm{val}}$ validation criteria in order to determine the optimal reconstruction. As discussed in the main text, we are interested in the effect of the modification of the simulation hyper-parameters on the the quality of the spectral reconstruction. 

As above, we focus in particular on the effective inverse temperature, $\Theta$, the number, $N_\omega$, of $\delta$-functions considered in the $\omega$-interval $[0, 5]$, and the uniform shift, $\delta\omega$. Note that, contrary to the case of the harmonic oscillator, the spectral function is now expressed in terms of a set of {\em non-equally-spaced} $\delta$-peaks (see Table~\ref{table::dwell frequencies}). As a  consequence, none of the nodes of the uniform $\omega$-grid we consider in our calculations coincides with the $\omega_i$. This is an intentionally non-optimal choice which, however, allows us to further illustrate important features of the reconstruction and associated validation, and provides hints towards possible improvements.

We first consider the inverse temperature $\beta=8$, and choose $\Delta\tau=0.1$. At this low temperature, all coefficients at frequencies larger than $\omega_1=E_1-E_0$ in Eq.(\ref{somega}) are suppressed exponentially, and the spectrum practically consists of a single $\delta$-peak. As for any discrete spectrum, the reconstruction becomes quite sensitive to the positions of the frequencies $\omega_i$ if the grid is  sparse. One can therefore study in detail the accuracy of the reconstruction when modifying the shift, $\delta\omega$, and the  distance between the grid points. Here, however, we do not perform such a detailed analysis, and for the sake of illustration we consider only  a few different models with a  regular grid shifted by a constant amount.

In Fig. \ref{fig::dwell spectrum beta=8} we show the resulting spectral functions at the indicated values of $N_\omega$ and $\delta\omega$, while  Fig.~\ref{fig::dwell chi valid b8} displays the corresponding $\chi^2_{\mathrm{val}}$. The data for $N_\omega=10$ and $\delta\omega=0$ correspond to values of $\chi^2_{\mathrm{val}}$ approximately ten times larger than those associated with the shifted lattice case, $\delta\omega=0.25$, and have been thus been omitted. If we consider the inaccuracy in the alignment of the $\omega$-grid with the exact peak positions, it comes as no surprise that the model with $N_\omega=10$ and $\delta=0.25$ provides a worse reconstruction than what we have obtained with the non-shifted $N_\omega=25$ grid, despite visually resembling more closely  the exact spectrum. 
Reconstruction of a single peak spectrum is performed essentially with only a few coefficients $A(\omega_i)$ that are closest to $\omega_1$, and because with denser grids we have more fitting parameters at our disposal,  it is natural that one obtains a better validation with $N_\omega=25$.

We now consider the more challenging case of the correlation function $C_{xx}^{\beta=1}$ at $\beta=1$ (calculated with $\Delta\tau=0.02$), whose spectrum displays several peaks (Fig.~\ref{fig::dwell spectrum beta=1}). We perform the reconstruction with the same parameters as in the case above.The use  of sparse grids (even inaccurately placed) proves advantageous in this case. Both models with $N_\omega$=10 perform considerably better than in the previous case, whilst the shifted lattice shows the best fit among considered models at $\Theta=10$ (Fig.\ref{fig::dwell chi valid b1}). This figure also clearly demonstrates the effect of the overfitting, which manifests itself in the plot as the increase of the validation $\chi^2_{\mathrm{val}}$ after some value of $\Theta$ for each configuration of parameters.

Another practical issue that is interesting to address is the influence of the imaginary time discretization of the test set of correlation function. Using sparser data set for calculation of $\chi^2_{\mathrm{val}}$ we do not observe any qualitative effect on the result. Value of $\Theta$, at which the best fit (minimum of $\chi^2_{\mathrm{val}}$) is achieved, however, becomes smaller with less test data.
}

\bibliography{references}
%
\end{document}